Between Copyright and Computer Science:

The Law and Ethics of Generative AI

By

Deven R. Desai and Mark Riedl

ABSTRACT

Copyright and computer science continue to intersect and clash, but they can coexist. The advent of new technologies such as digitization of visual and aural creations, sharing technologies, search engines, social media offerings, and more challenge copyright-based industries and reopen questions about the reach of copyright law. Breakthroughs in artificial intelligence research, especially Large Language Models that leverage copyrighted material as part of training models, are the latest examples of the ongoing tension between copyright and computer science. The exuberance, rush-to-market, and edge problem cases created by a few misguided companies now raises challenges to core legal doctrines and may shift Open Internet practices for the worse. That result does not have to be, and should not be, the outcome.

This Article shows that, contrary to some scholars' views, fair use law does not bless all ways that someone can gain access to copyrighted material even when the purpose is fair use. Nonetheless, the scientific need for more data to advance AI research means access to large book corpora and the Open Internet is vital for the future of that research. The copyright industry claims, however, that almost all uses of copyrighted material must be compensated, even for non-expressive uses. The Article's solution accepts that both sides need to change. It is one that forces the computer science world to discipline its behaviors and, in some cases, pay for copyrighted material. It also requires the copyright industry to abandon its belief that all uses must be compensated or restricted to uses sanctioned by the copyright industry. As part of this re-balancing, the Article addresses a problem that has grown out of this clash and is under theorized.

Legal doctrine and scholarship have not solved what happens if a company ignores Website code signals such as "robots.txt" and "do not train." In addition, companies such as the *New York Times* now use terms of service that assert you cannot use their copyrighted material to train software. Drawing the doctrine of fair access as part of fair use that indicates researchers may have to pay for access to books, we show that same logic indicates such restrictive signals and terms should not be held against fair uses of copyrighted material on the Open Internet.

In short, this Article rebalances the equilibrium between copyright and computer science for the age of AI.

Between Copyright and Computer Science:

The Law and Ethics of Generative AI

By

Deven R. Desai* and Mark Riedl†

* Sue and John Stantan Professor of Law and Ethics, Georgia Institute of Technology, Scheller College of Business, Associate Driector, Law, Policy, and Ethics Georgia Tech Machine Learning Center; J.D., Yale Law School; Affiliated Fellow, Yale Law Information Society Project; former Academic Research Counsel, Google, Inc. This Article was supported in part by summer research funding from the Scheller College of Business. We are grateful for the feedback from participants at the Legal Scholars Roundtable on Artificial Intelligence hosted by Emory Law School. The views expressed herein are those of the authors alone and do not necessarily reflect the view of those who helped with and supported this work.
† Professor Georgia Institute of Technology, College of Computing, School of Interactive Computing and Associate Director of the Georgia Tech Machine Learning Center; PhD in Computer Science, North Carolina State University.

# Introduction

Predictions from legal scholars that machine learning, the data machine learning needs, and copyright were headed for a clash with fair use have come to a head.[1] Large Language Models (LLMs) need huge amounts of data, and that data is often protected by copyright. LLMs are the fuel behind popular, revenue generating services from companies such as OpenAI, Anthropic, Cohere, and AI21 Labs, among others.[2] The fair use analysis around non-expressive use of copyrighted material—i.e., you can use copyrighted material to train software—has sound logic and appears strong.[3] Yet as scholars have noted, the advent of LLMs and what they can do has highlighted "uncertainties" around the doctrine such that the potential for extreme changes to the law were possible.[4] Aggressive deployment of LLM-driven "generative AI" services and products built on copyrighted material—including illicit book libraries—means those possible extreme changes are now upon us. Some options are severe enough to harm AI research and indeed change the nature of the Open Internet,[5] but that does not have to be the case. Put simply, the current lawsuits and proposed laws over applications, such as OpenAI's ChatGPT, threaten basic research in AI, miss important points about the technology behind the offerings, and yet raise strong questions about the law, ethics, and future of AI. As such, in this Article we look at the science and business practices around AI research and offer a way to maintain the balance between copyright and computer science.

---

[1] *See e.g.*, James Grimmelmann, *Copyright for Literate Robots*, 101 IOWA L. REV. 657 (2015); Benjamin LW Sobel. Artificial Intelligence's Fair Use Crisis. 41 COLUM. JL & ARTS, 45 (2017); Peter Henderson, Peter, Xuechen Li, Dan Jurafsky, Tatsunori Hashimoto, Mark A. Lemley, and Percy Liang *Foundation Models and Fair Use, arXiv preprint arXiv:2303.15715* (2023).

[2] *See e.g.,* Aaron Kantor, *Top 11 OpenAI Competitors & Alternatives Revealed*, THE BUSINESS DRIVE, July 21, 2023, at https://thebusinessdive.com/openai-competitors#openais-strongest-competitors

[3] Matthew Sag, *Fairness and Fair Use in AI*, FORDHAM L. REV. (forthcoming) at https://papers.ssrn.com/sol3/papers.cfm?abstract_id=4654875

[4] *See* Henderson et al., *supra* note 1.

[5] *See infra* Part III.A.2 (explaining the shift to new signals to prevent scraping the Internet and training software based on the data gathered) *Cf.* Pamela Samuelson, *Generative AI Meets Copyright* 381 SCIENCE 158, 159 (2023) (arguing use of "a dataset consisting of 5.85 billion hyperlinks that pair images and text descriptions from the open internet" likely lawful given the dataset was made by a European non-profit and protected by EU law on using copyrighted material for text and data mining).

OpenAI's ChatGPT has astonished the world and seems to be a leap towards artificial general intelligence (AGI) machines,[6] but it isn't.[7] Even OpenAI's CEO, Sam Altman, has admitted that the term is "ridiculous and meaningless."[8] Indeed, ChatGPT is only one offering of software based on research in natural language processing (NLP) and specifically LLMs.[9] LLMs leverage patterns in data to predict what might come next in a sentence, paragraph, or essay; sometimes with incredible fluidity and accuracy.[10] But these offerings are not magical AI.[11] They are services and products. Nonetheless, claims, or perhaps faith, that researchers are pursuing AGI matter, because people can forget the difference between how academic research makes progress on AI research and the reality of a non-academic group turning research into products and services.

---

[6] Sara Morrison, *What Microsoft Gets By Betting on the Maker of ChatGPT*, VOX (January 23, 2023) at https://www.vox.com/recode/2023/1/23/23567991/microsoft-open-ai-investment-chatgpt

[7] Mark Reidl, Toward AGI – What's Missing? MEDIUM (August 3, 2023) at https://mark-riedl.medium.com/toward-agi-what-is-missing-c2f0d878471a; Sean Illing, *Stuart Russell Wrote the Textbook on AI safety. He Explains How To Keep It from Spiraling Out of Control*, Vox (September 20, 2023) at https://www.vox.com/the-gray-area/23873348/stuart-russell-artificial-intelligence-chatgpt-the-gray-area

[8] Kevin Roose, *I Think We're Heading Toward the Best World Ever': An Interview With Sam Altman*, NY Times (November 23, 2023) at https://www.nytimes.com/2023/11/20/podcasts/hard-fork-sam-altman-transcript.html

[9] *See e.g.*, Dorothy Neufeld, *Visualizing the Training Costs of AI Models Over Time*, VISUAL CAPITALIST, June 4, 2024 (noting costs to develop LLM models for a range of companies).

[10] LLMs are also increasingly able to take images as input alongside text to fuel image generators. The technical and legal issues around images and generative systems are, however, different enough to be beyond the scope of this paper. So too for sound and music. *See e.g.*, Anna Washenko, *AI Startup Argues Scraping Every Song on the Internet Is Fair Use*, ENDGADGET, August 1, 2024 at https://www.engadget.com/ai/ai-startup-argues-scraping-every-song-on-the-internet-is-fair-use-233132459.html; Daniel Tencer, *As Suno and Udio Admit Training AI with Unlicensed Music, Recording Industry Says: "There's Nothing Fair About Stealing an Artist's Life Work"*, Music Business Worldwide, August 5, 2024 at https://www.musicbusinessworldwide.com/as-suno-and-udio-admit-training-ai-with-unlicensed-music-record-industry-says-theres-nothing-fair-about-stealing-an-artists-lifes-work/

[11] *Cf.*, Deven R. Desai and Johsua A. Kroll, *Trust But Verify*, 31 HARVARD J. OF L. AND TECH. 1, 4 (2017); *accord* Ian Bogost, *The Cathedral of Computation*, THE ATLANTIC (Jan. 15, 2015) (("The next time you hear someone talking about algorithms, replace the term with 'God' and ask yourself if the meaning changes. Our supposedly algorithmic culture is not a material phenomenon so much as a devotional one." http://www.theatlantic.com/technology/archive/2015/01/the-cathedral-of-computation/384300/

Although LLMs are not AGI, the quest for it, or what was originally called "genuine intelligence,"[12] animates the motives and methods behind LLM and other academic AI research. Understanding those motives and methods helps understand the legal and ethical implications of AI research in general as compared to commercial offerings based on AI research. In simplest terms, calls to limit or regulate AI research, such as to allow data to be scraped but not be used to train software or to charge for any use of copyrighted material to train software, are an over-correction that would harm future, desired AI research as called for by President Biden's recent executive order on AI. Nonetheless, AI researchers need to understand that just because they can do something, does not mean they should.[13]

Thus, this paper explains the theory and practice behind LLMs so people can recognize facts over fears and fictions and offers a plan to resolve the legal and ethical issues around the future of LLMs and AI research. Part I lays out what LLMs are and explains the technical and theoretical aspects of LLMs. It explains why vast amounts of data—such as book corpora and large crawls of the Internet—are vital for AI research. It also explains the technical aspects of LLMs including the relationship between generalization and memorization. Once you understand that relationship, you can understand the differences between LLM research seeking to emulate language and LLM research that wants to be a commercial answer-providing product. That understanding is necessary to unravel the legal issues around LLM research and products. Part II explains what happens when corporate researchers forget they are not in academia. The needs and practices of academic and commercial AI research often converge, but most importantly, they also diverge. This Part shows why cultural and practical limits around academic research limit potential harms and legal issues as opposed to non-academic research efforts, which use so many resources that commercialization and effects on markets are almost inevitable. In short, solutions to how computer science uses copyrighted material should accommodate academic needs and manage commercial issues, but current approaches fail to appreciate the difference between the two endeavors.

---

[12] MELANIE MITCHELL, ARTIFICIAL INTELLIGENCE: A GUIDE FOR THINKING HUMANS, 9 (2019)
[13] Ayana Howard and Deven R. Desai, *Taming AI's Can/Should Problem*, MIT SLOAN MANAGEMENT REVIEW, (May 18, 2021), at https://sloanreview.mit.edu/article/taming-ais-can-should-problem/

Part III accepts that all AI research is under scrutiny and presents a legal and ethical guide for gathering and using data for research. The Part focuses on legal issues particular to books and to Internet data. It explains that the better the breadth of either type of copyrighted data, the better an LLM can capture a diversity of voices and diversity of knowledge. Nonetheless, the Part takes a position contrary to some legal scholars and argues the assumption that you can use hundreds of thousands of illegally copied and obtained books rests on precarious grounds.[14] It adds to the literature by addressing evolving issues around expanded use of code signals--such as robots.txt, no crawl, and do not train--and restrictive terms of service. The part shows that these changes and uncertain law around what it means to ignore such restrictions threatens not only AI research but also signifies a shift from an Open Internet to a more closed, permission based one. The Part argues that we should not repeat the mistakes of letting the copyright industry dictate terms for computer science research, because that would create a new wave of litigation and what Professor Heller has explained is a Gridlock Economy.[15] The Part then shows how the law around platforms,

---

[14] *See infra*, Part III.A.1 *Cf.* Anupam Chander and Madhavi Sunder. *The Romance of the Public Domain*. 92 CALIF. L. REV. 1331 (2004) (examining the potential harms from over-stated claims about what should be in the public domain).

[15] *See* MICHAEL HELLER, GRIDLOCK ECONOMY: HOW TOO MUCH OWNERSHIP WRECKS MARKETS, STOPS INNOVATION, AND COSTS LIVES (2008) (detailing the way in which too many and fragmented property rights create an anticommons or wasteful underuse)

intellectual property harm mitigation, and mistakes in Silicon Valley research provide a set of practices that can help protect commercial entities that use copyrighted material to train their software.

Part IV presents solutions to maintain and expand access to books and to maintain an Open Internet. For books, we argue that if courts hold that you need a legitimate copy of a book to use it as a data source, courts should also rule that using tools to circumvent digital rights management is allowed when the end purpose if fair use. Nonetheless, asking all researchers to buy digital books or buy hard copies and then scan them, as Google did in its Google Books Project, is inefficient. As such, we offer a radical solution informed by our analysis of the contours of fair use and the Google Books and Hathi Trust cases. Society should offer a secure, training book repository for research. Given that libraries have already worked with Google and Hathi Trust and that the digital libraries are fair use under the courts' decisions, the libraries could work with either project to offer such a service at prices designed to enhance access and provide a share to authors rather than be a font of cash. Our solution leverages a point about the desired research. You do not have to give the data to the researcher. Instead, the repository could host the data securely and allow the researcher to train their model on the data. Rather than "shadow libraries" floating around the Internet, the data would stay at the repository. As this option cannot be required, Congress may need to pass a law setting up the rates and payments. Given that hurdle, we offer a simpler one based on the Audio Home Recording Act. There should be an added fee on book sales collected and distributed to authors, but only if the law also enshrines using book corpora even if they come from shadow libraries. To maintain an Open Internet, we offer that the logic of the Supreme Court's decision in *Campbell v. Acuff-Rose Music, Inc.*,[16] where one need not ask permission when the use is fair and the fact that access was legal should mean that unless Internet content is behind a paywall, data scraping for software training should be deemed fair use regardless of the code signals or terms of service put on such data. The paper then concludes.

[16] 510 U.S. 569, 585 n. 18 (1994).

## I. The Why and How Behind LLMs

AI research seeks large datasets because large datasets are a cornerstone not only of LLM research, but of modern AI research in general.[17] Calls to either ban or charge high prices for using copyrighted material such as book corpora, writing on the Web, and images on the Web threaten fundamental research in AI. Understanding the roots of artificial intelligence research explains why people pursue AI research, including LLMs, and why such research needs large datasets.

The attendees of the famous Dartmouth conference of 1956 were fascinated by the idea of creating a thinking machine and convinced that such a goal was attainable. One professor, John McCarthy, coined the term "artificial intelligence." No one liked it, in part because the goal was to create "genuine intelligence."[18] That goal raises a key, unanswered question: what is intelligence? The proposal for the first conference argued that "every aspect of learning or any other feature of intelligence can be so precisely described that a machine can be made to simulate it."[19] The idea sounds like a singular path would be taken, but even that initial workshop listed several possible paths forward. Attendees pursued a theory of "Neuron Nets" to emulate the way humans create concepts; "Self-Improvement" where a machine is "truly intelligent" if it learns on its own; "How Can a Computer Be Programmed to Use a Language" which investigated emulating the human thought as it operates within human use of language; and "Randomness and Creativity" the ability of a machine to do imagine new things with some randomness guided by intuition so that the outcome is efficient; among other topics as avenues to "genuine intelligence."[20] Then as now, there is no one path to AGI. But breakthroughs in one area can lead people to assert, if not believe, that AGI is here. The recent changes in how machines use language are a recent example of such exuberance.

---

[17] STUART RUSSEL AND PETER NORVIG, ARTIFICIAL INTELLIGENCE: A MODERN APPROACH 44 (4th Ed. 2020) (explaining the importance of "big data" and how it fuels modern AI research)
[18] MELANIE MITCHELL, ARTIFICIAL INTELLIGENCE: A GUIDE FOR THINKING HUMANS, 19 (2019)
[19] John McCarthy, Marvin L. Minsky, Nathaniel Rochester, and Claude E. Shannon, *A Proposal for the Dartmouth Summer Research Project on Artificial Intelligence, August 31, 1955*. 27 AI Magazine 12 (2006).

Language is considered a defining aspect of being human and thinking.[21] LLMs are part of Natural Language Processing (NLP), a core area of AI research that tries to emulate the way humans use language. If you can show that a machine can use language similar to how humans use language, you might claim you have taken an important step forward on one of the paths to machine intelligence set out at the Dartmouth conference. The evolution of NLP methods shows how we arrive at the LLMs of today. Because LLMs use techniques common across AI research, that understanding gives insights on the data and computational needs for much of AI research. Those needs go to the heart of questions about how that research is done. In short, the need for vast amounts of data is real and cutting off access to such data poses threats not only to commercial LLM research but to future AI research in general.

[20] John McCarthy, Marvin L. Minsky, Nathaniel Rochester, and Claude E. Shannon, *A Proposal for the Dartmouth Summer Research Project on Artificial Intelligence, August 31, 1955*. 27 AI Magazine 12 (2006).
[21] *See e.g.*, RUSSEL AND NORVIG, *supra* note 17, at 874-875 (noting that Alan Turing used language as his test for intelligence "because of its universal scope and captures so much of intelligent behavior); *accord* Jared Diamond, THE RISE AND FALL OF THE THIRD CHIMPANZEE (1991) ("After all, it is language that allows us to communicate with each other far more precisely than any animal can. Language enables us to formulate joint plans, to teach one another, and to learn from what others have experienced elsewhere or in the past.")

A. The Evolution of Natural Language Processing, A Vital Part of the Quest for AGI

Early NLP work relied on "symbolic rule-based approaches—that is, programs that were given grammatical and other linguistic rules and applied these rules to input sentences."[22] That approach fits within the idea that humans use rules and variables as we speak.[23] If you can capture those rules and variables, you can program a machine to use language the way a human does. This approach seems elegant because it implies that language is produced through enumerable procedures and does not require a huge amount of data. As with other areas of AI research that followed symbolic approaches, NLP systems were, however, found to be "brittle."[24] That is, regardless of how many rules about language the machine has, there will still be things that the machine cannot understand or produce the way humans do.[25] The symbolic rule-based approaches revealed that the way humans use language can often violate the formal rules of grammar but still be comprehensible by people.[26]

To address brittleness, NLP researchers in the late 1980s and 1990s turned to statistical approaches that capture the probabilistic relationship between words.[27] The idea, as counterintuitive as it might be, was that it didn't matter what words meant or what the grammatical relationship between words is, if they follow patterns.[28] This concept can be traced back to Claude

---

[22] Mitchell, *supra* note 18, at 180.
[23] *Id*.
[24] *Id*. at 198.
[25] ML researchers used to use "brittle" to attack the symbolic AI researchers. In theory learning systems wouldn't be as brittle, but were never good enough to really have to worry about it. Now, ML researchers are talking amongst themselves about how to overcome brittleness where "brittleness implies that the system functions well within some bounds and poorly outside of those bounds." Andrew J. Lohn, *Estimating the Brittleness of AI: Safety Integrity Levels and the Need for Testing Out-of-Distribution Performance*." arXiv preprint arXiv:2009.00802 (2020).
[26] *Id*.
[27] *Id*.
[28] Mitchell, *supra* note 18, at 181 (describing a system that captures language usage well but without "any understanding of the meaning").

Shannon's seminal work on how to define information.[29] According to Shannon's theory, "the informative content of a message is directly related to the degree of uncertainty regarding its content. Predictable messages have a low information content; unexpected or surprising messages have a high information content."[30] The focus shifts to what you can learn from the message.

When the message is expected, there is nothing new and so nothing to learn. If human writing has regularity, then it should be learnable and non-random. We capture this with the mathematical notation $P(w_n|w_1, w_2, \ldots, w_{n-1})$ which reads: the probability of the last word in a sequence (the n^th word) is dependent on the words that precede it from the first word to the second to last word. Intuitively this makes sense. When encountering the sequence of words "The deep blue", we expect the word "sea" to be highly likely to appear, but we do not expect the word "cucumber" to be likely to appear. This equation has a name: a *language model*.

A language model captures the statistical probability of a sequence of words—the context $(w_1, w_2, \ldots, w_{n-1})$—and its successor $(w_n)$.[31] The term was first introduced in 1976 but it wasn't until the late 1980s that computers became powerful enough to extract the large probability tables from raw text data. Early language models were simple.[32] The simplest is called the bigram model, which just probabilistically relates a word to the immediately preceding word: $P(w_n|w_{n-1})$.[33] If we imagine 50,000 words in the English language (an under-estimate), then we need a table with $50{,}000 \times 50{,}000$ numbers to store all the probabilities between pairs of words. To "learn" these probabilities, you must simply fill in this table by scanning through a corpus of human-written text and counting how often one word is preceded by another word.[34]

---

[29] C.E. Shannon, *A Mathematical Theory of Communication*, 27 BELL SYS. TECHNICAL J. 379, 379 (1948). Shannon's insights trace to Markov's work on n-grams being used to predict the next letter in a work by Pushkin. Daniel Jurafsky and James H. Martin, SPEECH AND LANGUAGE PROCESSING: AN INTRODUCTION TO NATURAL LANGUAGE PROCESSING, COMPUTATIONAL LINGUISTICS, AND SPEECH RECOGNITION 25 (2023)

[30] Dan Burk, *The Problem of Process in Biotechnology*, 43 HOUSTON L. REV. 561, 584 (2006)

[31] *See* Jurafsky and Martin, *supra* note 29 at 4 (explaining n-grams); STUART RUSSEL AND PETER NORVIG, ARTIFICIAL INTELLIGENCE: A MODERN APPROACH 877 (4TH ED. 2020) (same).

[32] Frederick Jelinek, *Continuous Speech Recognition by Statistical Methods* 64 PROCEEDINGS of the IEEE (1976); Jurafsky and Martin, *supra* note 29, at 25 (noting resurgence of n-gram research).

[33] Jurafsky and Martin, *supra* note 29 at 3-4 (explaining bigrams) at https://web.stanford.edu/~jurafsky/slp3/3.pdf; RUSSEL AND NORVIG, *supra* note 17, at 877.

Despite being computationally expensive, this simple bigram model can be used to implement the autocomplete functionality that we have on modern smart phones. Autocomplete gives a small sampling of highly probable words for you to choose from instead of typing a word from scratch. Nonetheless, being able to predict a word based only on the previous word has some limitations. If we only look at the previous word to predict the next one, "The deep blue" and "A sweet blue" are indistinguishable. As such there is a need to look beyond the previous word, which is what happened next in NLP research.

After the development of the bigram model, more complicated models have been developed such as the trigram, which gives the probability between a word and the two words immediately preceding it: $P(w_n|w_{n-2}, w_{n-1})$.[35] This model can capture the difference between "deep blue" and "sweet blue" to predict "sea" and "berry", respectively. This comes at greater computational cost, now requiring $50{,}000 \times 50{,}000 \times 50{,}000$ probability values. Autocomplete becomes a better guesser of words as the model goes from bigram to trigram and beyond to 4-gram and 5-gram.[36] Although no one would confuse 5-gram models with human writing, it works well enough that it is still used for many applications such as modern-day autocomplete functions that can predict the next five words in a phrase. The models are useful but if we are trying to reach a point where a machine can use language more like humans, the models come up short.[37] In simplified terms, the model "forgets" what was "said" more than five words ago. To move beyond that limit requires an ability to build models with more context which leads to the next step on our way to LLMs.

---

[34] The table doesn't need to be in any particular order. If you imagine a spreadsheet, you would have all the words down the side. These are the words in position one. Then across the top you would have all the words again. These are the words that might show up in position 2. The cells would contain numbers between 0 and 1 indicating the probability that the word in the column follows the word on the row. For an example of a bigram probability table *see* Jurafsky and Martin, *supra* note 29 at 5-6.

[35] *Id.* at 2-3. (explaining probability approach to NLP).

[36] STUART RUSSEL AND PETER RUSSEL AND NORVIG, *supra* note 17, at 883 (comparing how well different n-gram models work)

As a language model brings in more context, it is no longer practical or feasible to count words and build increasingly larger probability tables. The solution to this problem is to use a particular type of machine learning called artificial neural networks, which learn compact approximations of an underlying phenomenon. The amount of memory—or size of the neural network—can be traded off for accuracy. That is, as we allow the model to get larger, its approximation gets closer to the true underlying phenomenon. In the case of language modeling, the neural network approximates *word probabilities* instead of looking up *precise values* in a table.

Although there has been a considerable evolution of work on neural network language models from the late 1990s, the key breakthrough was a particular type of neural network called the *transformer*.[38] Transformers are possible in part because of the recent availability of very large cloud computing systems and very large amounts of easily attainable text on the Web, scanned books, and other sources. To be clear, the amount of data and computing power needed for transformers are big, resource intensive changes as compared to where NLP research started with a mere 50,000 x 50,000 word probability table. Transformers are at the core of LLMs, and we now turn to how they work.

---

[37] RUSSEL AND NORVIG, *supra* note 17, at 883 (noting "limits" to n-gram models))
[38] Ashish Vaswani, Noam Shazeer, Niki Parmar, Jakob Uszkoreit, Llion Jones, Aidan N. Gomez, Łukasz Kaiser, and Illia Polosukhin, *Attention Is All You Need* ADVANCES IN NEURAL INFORMATION PROCESSING SYSTEMS 30 (2017); *see also* RUSSEL AND NORVIG, *supra* note 17, at 883-884 (noting the way transformers improve NLP results as compared to n-grams).

## B. LLMs: The Next Big Leap for NLP

The transformer is so named because it incrementally transforms an input sequence of data, called tokens,[39] in a way that makes it easier to guess the next word in the stream.[40] It does this with a mathematical concept called self-attention.[41] While we will not detail how this works, we will convey the intuition about how and why it works.

### 1. Tokens Need Large Datasets and Computing Power

Consider a sentence such as "Sam rode his dirty bike home and". Each word is a token (sometimes longer words are broken down into several tokens). Each token is initially replaced by a unique string of numbers. For example, "dirty" might be <0.12, -1.7, 0.23, …, 1.55> with a total of 512 numbers, called an embedding. Through repeated practice, the transformer has learned that certain words go together—or "attend" to each other—in particular ways due to the relative regularity of human language (English in this case).[42] "Dirty" modifies "bike" but not "his". Attention puts words that attend to each other together, combining their numerical values in a way that can be metaphorically thought of as creating new words through the merging of words; "dirty" and "bike" becomes a word that might convey the concept of a bike that is dirty, "I" and "my" are combined, "rode" and "Sam" (the subject) and "bike" (the direct object) become a new concept meaning roughly that Sam is riding a bike.[43] The self-attention transformation is applied several times over, creating more and more complex concepts: it isn't just that Sam rode a bike but that they rode something that was conceptually a bike that was dirty.[44] As you might imagine, being able to make up specialized terms and concepts to mean complex combinations of verbs, entities, and relations means that by the time this process is done, the model has a much better shot at correctly guessing what might possibly come next.

---

[39] Jurafsky and Martin, *supra* note 29 at 17-23 (explaining tokenizing)
[40] Vaswani, et al., *supra* note 38, at 3; *see also* Russel and Norvig, *supra* note 17, at 919-922 (explaining self-attention on transformer architecture).
[41] *Id.*
[42] *See* Vaswani, et al., *supra* note 38, at 3 (defining attention); *see also* Russel and Norvig, *supra* note 17, at 916-917 (setting out mathematical definition of attention).

Because in this context words are just strings of numbers, the transformation and guessing processes involves a series of mathematical matrix multiplications involving, in the case of ChatGPT3.5, around 350 billion numbers that we call parameters.[45] These matrix values are multiplied against the word embeddings to produce values from 0 to 1 on a merger scale. Some of these matrix multiplications produce values close to 1—indicating the words have a high probability of attending to each other and so should be connected when trying to predict what value comes next—or values close to 0 indicating that the words have a low probability of attending to each other and should not be connected when predicting what should come next.

But how does the transformer know when to attend and when not to attend? Initially, the parameters are set randomly, and, as a consequence, words randomly attend to each other. This produces nonsense mergers that do not increase the ability of the model to guess the next word. If the model does not guess the right word, then we compute an error score. This error score is used to figure out how to tweak the parameters so that the next time the matrix calculations do a better job at figuring out which tokens attend to each other. We measure the error score using another mathematical concept called based on entropy, which is a measure of how surprised the transformer is by the correct next word.

---

[43] *See* RUSSEL AND NORVIG, *supra* note 17, at 917 (using "the front door is red" as an example of the phrase attention is generating).
[44] *Id.* at 919-922 (explaining self-attention and how it works in a transformer).

[45] Chude Emanuel, *GPT-3.5 and GPT 4.0 Comparison: Exploring the Developments in AI-Language Models*, MEDIUM, August 3, 2023 (noting GPT-3.5 has about "200 billion more parameters" than GPT-3's 175 billion parameters) at https://medium.com/@chudeemmanuel3/gpt-3-5-and-gpt-4-comparison-47d837de2226; *see also* Helen Toner, *What Are Generative AI, Large Language Models, and Foundation Models?*, CSET, n. 1, May 12, 2023 ("Parameters are the numbers inside an AI model that determine how an input (e.g. a chunk of prompt text) is converted into an output (e.g. the next word after the prompt).").

The goal is to have less surprise within the context of the software's purpose. How surprise is tamed is handled in two ways: generalization and memorization.[46] We now turn to the way generalization and memorization work.

2. Reducing Cross-Entropy: Generalization versus Memorization

There are two ways to reduce entropy: generalization and memorization. Generalization happens when the model learns patterns that can be applied to situations it has never encountered before.[47] For language, generalization means that the parameters of the model have been configured to capture some generalized concept about how words relate to each other. In English, verbs are often preceded by subjects—the do-er of the verb. People who talk about things being dirty often follow with talking about things getting washed. People who talk about things being dropped often talk about things falling. Roughly, generalization is a property that supports emulating the way humans use a range of words and put them together into meaningful sentences. This has important implications for those who wish to make scientific contributions about the emulation of human language composition because---as in the case of the Turing Test---a human-like text composition system may have to respond to novel situations.[48] For example, we might want a chatbot that can respond to a request to describe what would happen to a cat if it were to find itself on Pluto, which might not be something that the model encountered in its training data.

[46] *Cf.* Aparna Elangovan, Jiayuan He, and Karin Verspoor. *Memorization vs. Generalization: Quantifying Data Leakage in NLP Performance Evaluation*. arXiv preprint arXiv:2102.01818 (2021) ("In the context of machine learning models, effectiveness is typically determined by the model's ability to both memorize and generalize.")
[47] *Id.* ("The ability of a model to generalize relates to how well the model performs when it is applied on data that may be different from the data used to train the model.").

As such, there is a simple fix for text LLMs that are supposed to aid composition but may cough up possibly infringing outputs such as entire books or chapters of books. Do not use the whole book.[49] Instead, use a good sample of a book such as cutting every fifth page of a novel or other type of book and disallow memorization of large chunks of text.[50] Or you might skip every fifth paragraph of a text if smaller granularity was desired. These tactics would allow the LLM to learn the pattern of the language but not be able to return large reproduction of the underlying data. But generalization means the LLM isn't trying to be precise. If you want to address certain specific topics and be accurate about them, generalization can come up short.[51] Prompts for summaries of a specific book or other data don't flow from general language composition. That's where memorization comes in.

Memorization happens when the model cannot find a general pattern,[52] such as is the case with facts such as the birthplace of Alexander Hamilton. There is no general principle that could be used to guess the birthplace, so the only way to reduce entropy when trying to guess Hamilton's birthplace is to memorize the answer. Memorization is thus a valuable property of LLMs if one is attempting to commercialize a question-answering system that can respond to questions with accurate knowledge.

---

[48] *See* Alan M. Turing, *Computing Machinery and Intelligence,* LIX Mind 433 (1950).
[49] During the editing of this paper, the theoretical point we make regarding how much of a book, or by extension a given corpus, by reducing the amount of the data that is tokenized was tested and appears successful. *See* Abhimanyu Hans, et al. *Be like a Goldfish, Don't Memorize! Mitigating Memorization in Generative LLMs*. ARXIV *preprint arXiv:2406.10209* (2024) (finding one can randomly exclude data in the tokenization process and thus reduce verbatim and related copyright reproduction issues but maintain generalization performance).
[50] *Id*.
[51] Michael Tänzer, Sebastian Ruder, and Marek Rei. *Memorisation versus Generalisation in Pre-trained Language Models*, 1 PROCEEDINGS OF THE 60TH ANNUAL MEETING OF THE ASSOCIATION FOR COMPUTATIONAL LINGUISTICS 7564, 7564 (2022) ("excellent generalization properties come at the cost of poor performance in few-shot scenarios with extreme class imbalances. Our experiments show that BERT is not able to learn from individual examples and may never predict a particular label until the number of training instances passes a critical threshold.").
[52] *Cf.* Elangovan et al., *supra* note 46 ("When there is considerable overlap in the training and test data for a task, models that memorize more effectively than they generalize may benefit from the structure of the evaluation data,

In a way memorization is not about prediction; it is about exact reproduction of passages within the training data.[53] Memorization means that some parameters in the model have been set to only work with a very specific sequence of input tokens and produce a very specific output token. For example, imagine you are introducing a professor at a conference, and you want to generate their bio. The professor works at University of Big State. There is no general property of the universe that would predict what university they work at, so to predict a token like "University" followed by "Big" followed by "State" as they attend to the professor correctly and thus reduce entropy, it would need to have learned a very specific pattern that only applies to a very small set of possible inputs. Put differently, that specific data are part of Long Tail data that is not easily predicted, but necessary for greater accuracy.[54] An LLM like ChatGPT will have to memorize the professor's bio if the system is to generate the bio accurately.[55] But it is a mistake to think generalization and memorization are siloed aspects of LLMs.

with their performance inflated relative to models that are more robust in generalization. However, such models may make poor quality predictions outside of the shared task setting.")

[53] *Cf.* id. ("When there is considerable overlap in the training and test data for a task, models that memorize more effectively than they generalize may benefit from the structure of the evaluation data, with their performance inflated relative to models that are more robust in generalization. However, such models may make poor quality predictions outside of the shared task setting.")

[54] Vitaly Feldman and Chiyuan Zhang, *What Neural Networks Memorize and Why: Discovering the Long Tail via Influence Estimation*, 33 ADVANCES IN NEURAL INFORMATION PROCESSING SYSTEMS 2881 (2020).

[55] This issue may be more acute for image generators. *Cf.* Matthew Sag *Copyright Safety for Generative AI*, __ HOUSTON L. REV. __ (forthcoming). If one wants to create an image with the Mona Lisa or other iconic image, the system will have likely to have the iconic image memorized to create the new image.

Generalization and memorization work hand-in-hand. Generalization may be used to compose the more formulaic parts of an answer to a question that contains a memorized fact, and memorization of facts about Pluto help in the application of commonsense involving cats in cold environments.[56] The tradeoff between generalization and memorization reveals that you have to know what your goal for your software is, if you want to assess whether it worked, and to avoid potential legal issues.

3. Does Your LLM Work? Better Yet, What's Your Goal?

In NLP research there are different, specific tasks with specific tests to see whether progress is made.[57] Tests include ways to assess common sense reasoning, question answering, textual entailment, and the ability to use language "in a way that is not exclusively tailored to any one specific task or dataset."[58] Whether your system generalizes, memorizes, or both affects how well the system performs on these tasks.

---

[56] *See* Tänzer, et al., *supra* note 51 at 7564 ("For many applications it is important for the model to generalise—to learn the common patterns in the task while discarding irrelevant noise and outliers. However, rejecting everything that occurs infrequently is not a reliable learning strategy and in many low-resource scenarios memorization can be crucial to performing well on a task"); *Cf.* Elangovan et al., *supra* note 46 ("An effective combination of memorization and generalization can be achieved where a model selectively memorizes only those aspects or features that matter in solving a target objective given an input, allowing it to generalize better and to be less susceptible to noise.")

[57] See e.g. Alec Radford, Karthik Narasimhan, Tim Salimans, and Ilya Sutskever, *Improving Language Understanding by Generative Pre-training*. (2018) (asserting GPT-1 performance on "natural language inference, question answering, semantic similarity, and text classification ... outperforms discriminatively trained models that employ architectures specifically crafted for each task, significantly improving upon the state of the art in 9 out of the 12 tasks studied.)

[58] *Id*. ("For instance, we achieve absolute improvements of 8.9% on commonsense reasoning (Stories Cloze Test), 5.7% on question answering (RACE), 1.5% on textual entailment (MultiNLI) and 5.5% on the recently introduced GLUE multi-task benchmark.")

The combination of generalization and memorization is why modern LLMs are especially good at following directions—the prompt may include clues as to what the user wants to see generated, which we might think of as instructions. That is, the more the prompt signals a pattern that triggers generalized and/or memorized parameters, the more an LLM is able to deliver plausible results. For example, if you want a scholarly essay about the Lincoln-Douglas debates, you need to account for the probability that the model was likely trained on a large number of example homework assignments available online. The prompt "Write an essay about the views in the Lincoln-Douglas Debates" may produce an output that notes this rough rule: "The essay should be five pages and cite sources," because the LLM's training data has a lot of examples of homework assignments and test questions. However, if the original prompt is followed by "suitable for scholarly publication," you are more likely to get an answer to the prompt versus an elaboration of possible assignment instructions.[59] But what if you are looking for something even more specific?

Suppose you want an essay about the reasons the North entered the Civil War and how the evolution of Lincoln's views on slavery influenced the decision to go war. An LLM may have been trained on an excellent scholarly article or book chapter on exactly these ideas. As opposed to a search query where a system will reach out to the web to find a source that seems to match your query and point you to that source, the LLM may find the prompt fits so well with the essay or book chapter, that the LLM returns all, or most, of the source. As a test of a system that can "converse" with the user as a human, the LLM has done well. Indeed, it will have done so well that it exceeds most people on the planet including many well-educated people unless they are quite well-read or experts on the Civil War. Yet, in this situation we have moved from prediction to something else.

[59] Most LLMs have been fine-tuned to prefer to try to answer a prompt as if it is a question or a request for information. This is called "instruction tuning". Shengyu Zhang, Linfeng Dong, Xiaoya Li, Sen Zhang, Xiaofei Sun, Shuhe Wang, Jiwei Li et al. *Instruction Tuning for Large Language Models: A Survey. arXiv preprint arXiv:2308.10792* (2023).

With enough data your LLM may be memorizing and giving memorized answers that fit a rather precise text rather than predicting what comes next. In that sense there is a paradox. Your system is not really predicting language outcomes or generating creative outputs. Instead, it is coughing up data. A scientist might care because the elusive idea of Artificial General Intelligence is in fact further away than one hoped. A commercial lab may have to admit that it is not near AGI and that the idea is silly.[60] These LLMs are not necessarily predicting language, let alone being a version of human thought as hoped or hyped by some in the field. Such LLMs offer commercial upsides even though they are a far cry from being AGI.

Indeed, Microsoft's initial use of Open AI's software was consumer-focused as part of Microsoft's search engine, Bing.[61] The goal was to displace Google's search engine dominance because a ChatGPT powered Bing would be able "to formulate and write out answers to questions instead of just putting out a series of links."[62] That result may be something consumers like, but it is essentially a somewhat richer way to present answers to a search query.[63] As further evidence of commercial goals of LLM companies, during the editing of this paper, OpenAI announced SearchGPT, OpenAI's foray into the lucrative search space,[64] and Perplexity has begun an ad revenue sharing model with news publishers when users' search queries (aka prompts) generate results that cite news sources.[65] Whether LLM-driven results are accurate, or violate current customs and laws are separate issues. More generally, the methods and goals for academic output as opposed to commercial output are quite different in ways that matter for AI research and regulation going forward.

---

[60] *See supra* note 8.
[61] *See supra* note 6.
[62] *Id*.
[63] Matteo Wong, *The AI Search War Has Begun*, THE ATLANTIC, July 30, 2024, (describing "AI powered search bar") at https://www.theatlantic.com/technology/archive/2024/07/perplexity-ai-search-media-partners/679294/
[64] Kylie Robinson, *OpenAI Announces SearchGPT, Its AI-powered Search Engine*, THE VERGE, July 25, 2024 at https://www.theverge.com/2024/7/25/24205701/openai-searchgpt-ai-search-engine-google-perplexity-rival

## II. The Difference Between Academic and Commercial Research

The contexts and goals of building a given LLM and/or the contexts and goals of that LLM's outputs highlight the tension between academic research and commercial *application* of research. An academic goal may be to emulate human use of language. A commercial goal may be to offer a service that users enjoy and for which they are willing to pay. Those differences matter when considering what rules should govern how we build AI software and the implications of, if not the liability for, outputs.

Corporate practices that ignore the realities of ethical and legal constraints do so at the peril of inviting lawsuits and ill-advised regulation. The problem is that regulations and other actions can affect both ill-conceived corporate acts and desired academic work. This section explains the differences between academic and corporate research to illustrate why recent reactions to corporate missteps should be tempered to accommodate academic work.

[65] Rebecca Bellan, *Perplexity Details Plan To Share Ad Revenue with Outlets Cited by Its AI Chatbot*, TECHCRUNCH, July 30, 2024, at https://techcrunch.com/2024/07/30/perplexitys-plan-to-share-ad-revenue-with-outlets-cited-by-its-ai-chatbot/

## A. Academic Research and Its Goals

Imagine it is 2015 and you are at a university working to advance NLP science. In simplified terms you need data, code to create and train a model, software that implements the system that delivers an intended purpose (be it chatbot, or something else) by running inputs through the model and processing the outputs. But before you can run the system software, you need computing power to end up with a trained final model in the first place. The amounts of data and computing power may start small but often end up growing. Roughly speaking, if you have more data in the dataset, and the model's increase in size is commensurate, the better the model will perform; and by more we mean a rather huge amount of data.[66] As your dataset grows so will your computing power needs. To start, you will choose data and that dataset may be relatively small as was the case with OpenAI's GPT-1 which used roughly 7,000 books in BookCorpus and had 117 million parameters.[67]

Assuming you have the data and computing power, you will run your experiments and publish a paper. That paper will provide ways to test your results. You will either share your data and your code to create the model, and/or the fully-trained model so the academic community can verify your results through replication. Unlike commercial research, in a way your work is done once your paper is published and the results are verified. The next step is to see what you or someone else can do to push the science and outcomes further by improving on your software. You will probably need more data and by extension more computing power.

---

[66] ChaptGPT-4 is now at 1.5 trillion parameters. *See* Emanuel, *supra* note 45 ("With 1.5 trillion parameters compared to GPT-3.5's 175 billion, GPT-4 clearly outperforms GPT-3.5 in terms of model size and parameters "). When ChatGPT-1 was released in June of 2018, it had 117 million parameters. ChatGPT-2, released in February 2019, had 1.5 billion parameters based on 8 million pages of web text. ChatGPT-3, released in June 2020, had 175 billion parameters. The number of parameters for each release jumped by more than a factor of ten which means the scale and complexity of the systems has grown at a tremendous rate. *See Bernard Marr, A Short History of ChatGPT: How We Got Where We Are Today*, FORBES (May 19, 2023) at https://www.forbes.com/sites/bernardmarr/2023/05/19/a-short-history-of-chatgpt-how-we-got-to-where-we-are-today/?sh=49d3213f674f

[67] *See supra* 57 at 4.

Data is cheap; computing is expensive. If you want a larger dataset, you will likely scrape the Internet, use a resource like CommonCrawl, and/or add in almost any other source of data you can. Once the data is gathered it needs to be tokenized and run through a transformer[68] to see how well it does in doing something, but that something can vary. Your goal could be predicting what comes next in a string of text larger than a phrase; perhaps shooting for a sentence, a paragraph, or even a short essay.[69] Your goal may be to see whether the system can interact with humans in ways that make a human think the system is in fact human.[70] The research will likely involve running experiments. It may take 10, 50, or even 100 iterations on the design of the new model to find the one that works as well as desired.[71] Those computations are expensive.[72] Your work may stop, because the practical reality is that experiments that cost millions of dollars are not likely to be done at an academic research institution.[73]

---

[68] Transformers are the core technology behind recent breakthroughs in LLMs. *See supra* note 38 and accompanying text.

[69] For example, within NLP there are different, specific tasks with specific tests to see whether progress has been made. *See e.g.* Radford, et al., *supra* 57, (asserting GPT-1 performance on "natural language inference, question answering, semantic similarity, and text classification ... outperforms discriminatively trained models that employ architectures specifically crafted for each task, significantly improving upon the state of the art in 9 out of the 12 tasks studied. For instance, we achieve absolute improvements of 8.9% on commonsense reasoning (Stories Cloze Test), 5.7% on question answering (RACE), 1.5% on textual entailment (MultiNLI) and 5.5% on the recently introduced GLUE multi-task benchmark.")

[70] This idea flows from Alan Turing's work. *See* Alan M. Turing, *Computing Machinery and Intelligence,* LIX Mind 433 (1950).

[71] Guido Appenzeller, Matt Bornstein, and Martin Casado, *Navigating the High Cost of AI Compute*, ANDREESSEN HOROWITZ, April 27, 2023 (estimating $560,000 for a single training run and noting "Multiple runs will likely be required.")

[72] *Id.*

[73] *Cf.* id. ("Training top-of-the-line models remains expensive, but within reach of a well-funded start-up. … generative AI requires massive investments in AI infrastructure today. There is no reason to believe that this will change in the near future. Training a model like GPT-3 is one of the most computationally intensive tasks mankind has ever undertaken."); *see also* Neufeld, *supra* note 9 (noting the foundational model, the transformer cost $930,000 in 2017; GPT3, $4.3 million in 2020; GPT-4 $78.4 million in 2023; and Gemini Ultra $191.4 million in 2023).

Put simply, the nature of academic research means there is no need to seek a monetary return on investment. The research is not usually funded with for-profit strings attached. As such, the model need not be deployed at scale with millions of users paying to use it or where issues around misuse or undesired biased outputs might affect a large part of the population. Things are different in commercial settings.

## B. Commercial Research and Its Goals

Sooner or later non-academic research that needs the best human talent, ever larger data sets and related, expensive computing power will have to pay for those resources. Imagine it is 2015 and to tackle the problem of funding needed for NLP research, some folks start a private, non-profit research group. As noted above, the first stage of research could be done in many academic research labs. The next stage is where problems occur. It is easy to underestimate the effect of these costs. OpenAI's early cloud computing costs were close to $8 million which was about 25% of its expenses.[74] Those resources allowed OpenAI to take the next step in its GPT development and pursue GPT-2. That work used BookCorpus, added 8 million web pages, and had 1.5 billion parameters, as opposed to the 117 million in GPT-1.[75] The results were another step forward in that "The diversity of tasks the model is able to perform in a zero-shot setting suggests that high-capacity models trained to maximize the likelihood of a sufficiently varied text corpus begin to learn how to perform a surprising amount of tasks without the need for explicit

---

[74] Stephen Nellis, *Microsoft to Invest $1 Billion in OpenAI*, REUTERS, July 22, 2019, at https://www.reuters.com/article/us-microsoft-openai/microsoft-to-invest-1-billion-in-openai-idUSKCN1UH1H9/

[75] *See Bernard Marr, A Short History of ChatGPT: How We Got Where We Are Today*, FORBES (May 19, 2023) at https://www.forbes.com/sites/bernardmarr/2023/05/19/a-short-history-of-chatgpt-how-we-got-to-where-we-are-today/?sh=49d3213f674f

supervision."[76] More simply, as OpenAI released GPT-2, people found it was better than expected at creating news stories, poems, answering questions, language translations, and other language tasks.

Once GPT-2 was released in early 2019, the interest in OpenAI's approach and how much it could be improved went up; as did the need for more resources. The research question didn't change much; the amount of data and computing power needed increased again. And a new need arose; the need to hire talent that might go to Google, Meta, and other corporations in the AI space. In short, OpenAI needed more money than non-profits usually attract. By one estimate OpenAI lost around $540 million in 2022.[77] To address the need, OpenAI created a system that was supposed to keep its non-profit corporation in charge but that had a for-profit company operating under the non-profit. That structure allowed OpenAI to offer equity to its employees and pursue partnerships such as its one with Microsoft. Microsoft invested $1 billion in OpenAI.

Unlike the academic world where return on investment is not an issue, OpenAI publishing scientific papers and receiving accolades for that work was not enough. The anticipated costs to build GPT-3 with its 175 billion parameters, the promise of equity returns to employees of the for-profit, and investment oversight from Microsoft meant money had to be generated. Shortly after Microsoft's investment, OpenAI announced it was pursuing ways to license its technology for profit as part of having funds to achieve AGI.[78] The need for returns has only increased given Microsoft's estimated $13 billion investment to date.[79]

[76] Alec Radford, Alec, Jeffrey Wu, Rewon Child, David Luan, Dario Amodei, and Ilya Sutskever, *Language Models Are Unsupervised Multitask Learners* February 14, 2019 available at https://cdn.openai.com/better-language-models/language_models_are_unsupervised_multitask_learners.pdf

[77] Hasan Chowdhury, *ChatGPT Cost a Fortune To Make with OpenAI's Losses Growing to $540 Million Last Year, Report Says*, BUSINESS INSIDER, May 5, 2023 at https://www.yahoo.com/tech/chatgpt-cost-bomb-openais-losses-125101043.html

Once OpenAI switched to commercialization the difference between lab research and commercial products became acute. A core problem was releasing versions of the software and the API to the world to see what would happen. Rather than worrying about "harm[ing] humanity and "unduly concentrating power" as OpenAI's Charter extolls,[80] the company threw the model online to see what would happen. As OpenAI's CEO, Sam Altman said, "What I think you can't do in the lab is understand how technology and society are going to co-evolve. …you just have to see what people are doing — how they're using it."[81] That approach reveals that the company either had not thought about its goals and the repercussions of releasing its software widely; didn't care; or as of its GPT-3 release was more interested in commercialization and the fees OpenAI charged for using its product. Recent revenue numbers—$1.6 billion annualized rate at the end of 2023 and $3.4 billion as of summer 2024—suggest commerce has been the driving force for some time.[82] To be clear, other players in this market such as Anthropic are also pursuing revenue and must do so to justify the investments they have received.[83] Even if LLM companies are saying they have built a sort of Swiss Army knife of software that can be used for a wide range of purposes, the exact use matters when it comes to understanding and mitigating potential harms and legal liabilities.

---

[78] *See* Jonathan Vanian, *OpenAI Will Need More Capital Than Any Non-Profit Has Ever Raised*, Fortune (October 3, 2019) ("OpenAI CTO Brockman said that the group's plan to develop artificial general intelligence (AGI) is "actually a really expensive endeavor" because of the enormous amount of computing resources required") at https://fortune.com/2019/10/03/openai-will-need-more-capital-than-any-non-profit-has-ever-raised/
[79] Jordan Novet, *Microsoft's $13 Billion Bet on OpenAI Carries Huge Potential Along with Plenty of Uncertainty*, CNBC (April 9, 2023) at https://www.cnbc.com/2023/04/08/microsofts-complex-bet-on-openai-brings-potential-and-uncertainty.html
[80] https://openai.com/charter
[81] *See supra* note 8.
[82] *See* Laura Bratton, *OpenAI's Revenue Is Skyrocketing*, QUARTZ, June 13, 2024 at https://qz.com/sam-altman-openai-annualized-revenue-triples-skyrockets-1851538234
[83] Reuters, *Anthropic Forecasts More Than $850 mln in Annualized Revenue Rate by 2024-end-report*, REUTERS, December 26, 2023 (noting investments by Amazon and Google and increased revenue projections for 2024) at https://www.reuters.com/technology/anthropic-forecasts-more-than-850-mln-annualized-revenue-rate-by-2024-end-report-2023-12-26/

Put differently, corporate AI researchers are often trained in, or have a history at, academic settings, but once in a corporate environment cultures and priorities change. Corporate research is far less restricted in resources and more restricted in outcomes. For example, both academic and corporate software research is data and computing power hungry. Corporate research, however, can access vast computing power, data, and money if it pursues outcomes aimed at commercial uses and products.[84] Afterall, a corporation that invests hundreds of millions, if not billions, in research expects returns, not Nobel Prizes or Turing Awards. No matter what a company says, once corporate money with corporate strings attached is taken, the ability to follow scientific ethics as opposed to classic corporate ethics diminishes and perhaps vanishes. Nonetheless, the next section offers a set of questions to guide AI research in both academic and commercial settings going forward.

## III. A Guide for Data in LLM Research

Regardless of academic or commercial contexts, AI research is under scrutiny, and AI researchers need to understand the ethical and legal boundaries of what they are doing. Although those boundaries may change depending on the context, we offer a set of questions that can guide research in both contexts.

---

[84] Guido Appenzeller, Matt Bornstein, and Martin Casado, *Navigating the High Cost of AI Compute*, ANDREESSEN HOROWITZ, April 27, 2023 (noting extreme expense to train models but that such costs are manageable for a "well funded start up").

## A. What Are My Data Inputs?

AI research must pay more attention to how it gathers data. The recent outcry over commercial generative software, such as OpenAI and other vendors have rushed to market, has put new pressure on the rules around what is allowed when gathering data to build an AI model.[85] Much of the furor flows from authors' lawsuits over the use of copyrighted material to build the models *and* the systems' early outputs. As people found that some LLM software coughed up results that might infringe copyright law--such as several pages of a novel or visual artwork that treaded close to infringing, existing copyrighted visual art--they were outraged about how LLMs are built. Even before the focus on LLMs, there have been ethical questions about whether to honor the robots.txt code[86] and legal questions about using unauthorized data from "shadow web libraries."[87] But given copyright's fair use doctrine and arguments that even access to illegal copies can be fair use, researchers tend to gather their data in this way.[88] This approach has been customary and accepted as fair use until OpenAI's recent reckless behavior.[89] That behavior has spawned a reaction that creates ethical, if not legal problems, for AI researchers.[90]

---

[85] *See e.g.*, Winston Cho, *New AI Lawsuit From Authors Against Anthropic Targets Growing Licensing Market for Copyrighted Content*, The Hollywood Reporter, August 20, 2024 (reporting about a lawsuit against Anthropic arguing that the company should have licensed book corpus material rather than used unauthorized material).
[86] *See* Mark Graham, *Robots.txt meant for search engines don't work well for web archives*, April 27, 2017 at https://blog.archive.org/2017/04/17/robots-txt-meant-for-search-engines-dont-work-well-for-web-archives/; Brad Jones, *Internet Archive will ignore robots.txt files to keep historical record accurate*, April 24, 2017 at https://www.digitaltrends.com/computing/internet-archive-robots-txt/#ixzz4gQYOqpUi
[87] *See* Sag, supra note *Fairness and Fair Use in AI*, FORDHAM L. REV. (forthcoming) at https://papers.ssrn.com/sol3/papers.cfm?abstract_id=4654875
[88] *See* Sag, supra note *Fairness and Fair Use in AI*, FORDHAM L. REV. (forthcoming) at https://papers.ssrn.com/sol3/papers.cfm?abstract_id=4654875
[89] *See* David Pierce, *The Text File That Runs the Internet*, THE VERGE, (February 14, 2024) at https://www.theverge.com/24067997/robots-txt-ai-text-file-web-crawlers-spiders
[90] *Id.*

The first step in any AI research is to gather data. Because models tend to perform better with more data, researchers will look for larger data sets. In LLM research the method itself mandates larger data sets. Recall that the shift from symbolic approaches to statistical approaches such as bigrams and 5-grams demand a decent data set that was computationally expensive.[91] Yet that approach fell short of using language the way humans do. The next step of using machine learning and artificial neural networks needs even more data and computational power. The shift to transformers requires even more data. This need is not simply a hunger for so-called "big data." The models are looking for large contexts to better predict what words come next. Thus, ChatGPT's evolution in size is related to how well it functions, and indeed each larger model performed better than the previous in significant ways. But from where does the data come?

How we choose what data to include creates an ethics paradox. More data means better performing models, but the sources of that data may be covered by copyright. The over-breadth of copyright protection means that not only books but also any writing, including writing posted to the Internet, is protected by copyright. In addition, online sources may have explicit, coded requests not to crawl the site or use the material for software training. Objections to using copyrighted material can range from absolutist—never use my writing to train software—to economic—you should pay to use my material.[92] We explore the differences between books and the Web to highlight the issues around using copyrighted material as software inputs.

[91] *See supra* notes 32to 38 and accompanying text.

[92] *See e.g.*, Winston Cho, *Authors Guild Exploring Blanket License For Artificial Intelligence Companies*, THE HOLLYWOOD REPORTER, January 11, 2024 (describing the Authors Guild's desire to create an opt-in licensing system for using books to train models).

1. Books as LLM Inputs

Legal precedent from the Google Books and Hathi Trust projects establishes that a commercial entity can copy copyrighted material—even millions of books—to index the material and enable metanalysis.[93] Even before that case, copyright law has allowed copying without permission to reverse engineer software, build software to detect plagiarism, and roughly "non-expressive uses of copyrighted works...as fair use."[94] As recent legal scholarship has explained, the problem comes from the outputs of such training; not the training itself.[95] We agree with that analysis. Nonetheless, OpenAI and other companies' reckless behaviors in releasing generative AI such as ChatGPT have called customary and accepted fair use practices into question. These reactions offer a chance to explain why data is needed and where choices in data gathering may matter more for future software development.

Let's assume you want to use only public domain material, that is, material not under copyright. Under U.S. law in 2024 your safest bet is anything published before 1929, because the copyright will have expired, and the work is in the public domain.[96] For anything published after 1929, you would have to track down whether authors obeyed copyright formalities such as registering the work, publishing with a copyright notice, and renewing the copyright on time with extra nuances depending on the dates when the work was published.[97] You'd spend a good amount of time and money trying to track down what is and isn't under copyright.[98] If you stick with public domain material--even with far less data than you'd have with scraping the Internet--you might be happy about avoiding potential copyright issues, having lower costs, and think, "Problem solved. I have enough data and it's free!" That position, however, runs into another problem.

---

[93] Authors Guild, Inc. v. HathiTrust, 755 F.3d 87 (2d Cir. 2014); Authors Guild v. Google, Inc., 804 F.3d 202, 216–17 (2d Cir. 2015).
[94] *See* supra note 55.
[95] *See e.g.,* Henderson et al., *supra* note 1.
[96] https://guides.library.cornell.edu/copyright/publicdomain
[97] *Id.*

A general, major critique of AI systems is that they *lack representative or inclusive data*.[99] Work published and available before 1929 is likely to be a less diverse set of authors, because diverse voices were not well-published, if at all, until recently.[100] Diverse here should include the whole range of authorship from all corners of humanity.[101] The paradox is that by respecting copyright, you would necessarily cut out the very voices that have said they are underrepresented in machine learning.[102] Diverse also should include a broad range of topics and inquiry as captured

---

[98] As scholars noted years ago, figuring out whether a book is under copyright is part of the orphan works issue which occurs when copyright holders "cannot be located by a reasonably diligent search." David R. Hansen, Kathryn Hashimoto, Gwen Hinze, and Pamela Samuelson *Solving the Orphan Works Problem for the United States*. 37 Colum. JL & Arts 1 (2013); Jennifer M. Urban, *How Fair Use Can Help Solve the Orphan Works Problem,* 27 BERKELEY TECH. L.J. 1379 (2012).

[99] *Cf.* Executive Order, 14110, Safe, Secure, and Trustworthy Development and Use of Artificial Intelligence , November 1, 2023 (explaining "bias" and "discrimination" have occured in AI systems and calling for prevention of future potential harm); Deven R. Desai, Swati Gupta, and Jad Salem, *Using Algorithms to Tame Discrimination: A Path to Diversity, Equity, and Inclusion* 56 UC DAVIS L. REV. 1703, 1714 n. 50, 1716 (2022) (examining claims that data necessarily has disparate impact or under-represents a group.

[100] *Cf.* Amanda Levendowski, How Copyright Law Can Fix Artificial Intelligence's Implicit Bias Problem, 93 WASH L. REV. 579 (2018); *accord* Sag, *supra* note 55.

[101] *Cf.* Danielle Allen, *We've Lost Our Way on Campus. Here's How We Can Find Our Way Back*, WASHINGTON POST (December 13, 2023) ("By diversity, we mean simply social heterogeneity, the idea that a given community has a membership deriving from plural backgrounds, experiences, and identities. Race, ethnicity, gender identity, sexual orientation, socioeconomic background, disability, religion, political outlook, nationality, citizenship, and other forms of formal status have all been among the backgrounds, experiences, and identities to which the Task Force has given special attention, but we have also attended to issues of language, differences in prior educational background, veteran status, and even differences in research methodologies and styles.") at https://www.washingtonpost.com/opinions/2023/12/10/antisemitism-campus-culture-harvard-penn-mit-hearing-path-forward/

[102] This problem is inherent to data of all kinds. You have to include individuals' data to respect their individuality. FRANK WEBSTER, THEORIES OF INFORMATION SOCIETY 55 (1995) ("[I]f we as a society are going to respect and support the individuality of members, then a requisite may be that we know a great deal about them."); *accord*

by the growth of publishing and expansion of scientific knowledge since 1929. Imagine if your dataset was stuck with nonfiction and science from 1929. The data will be quite biased towards an understanding of the world that is literally out-of-date and miss for example, the theory of chemical bonds, results from using the electron microscope, the discovery of nuclear fission, the science behind improving Mexican wheat and the resulting green revolution, the invention of the transistor, the discovery of DNA, even if we stop our list at around 1950.[103] Thus, having a larger book data set is good for your model because more data likely improves the model's performance on a pure predictive performance metric *and* your model works well for a larger range of voices. How to get the book data is, however, not yet solved.

Whether your research is at an academic or commercial entity matters as you collect book data. Matthew Sag argues that lawful access should not be required as a per se rule for non-expressive uses of copyrighted material even for commercial defendants.[104] That view means that if you are using copyrighted material to train the software, the use should be allowed. In that spirit, he argues that if a company such as OpenAI could not obtain digital copies of text "without a contractual promise not to engage in non-expressive use, faulting them for obtaining a copy in the shadowy corners of the Internet might seem a bit churlish."[105] This position raises at least two issues.

First, you are not required to ask permission to use IP *if* the use falls under fair use.[106] As the Supreme Court has pointed out, "Even if good faith were central to fair use ... being denied

---

Deven R. Desai, *Exploration and Exploitation: An Essay on (Machine) Learning, Algorithms, and Information Provision*, 47 LOY. U. CHI. L.J. 541, 577 (2015) permission to use a work does not weigh against a finding of fair use" because you might ask for permission "in an effort to avoid litigation."[107] Yet if you asked and were told "only use the data for non-expressive use," and you agreed to the terms, you could not use the material for an LLM at all if the purpose will be to create outputs that are expressive but not necessarily infringing on someone's copyright.[108] As such, you would be more likely to reject the terms and proceed with your offering based on the copyrighted work. Nonetheless, whether your use will be allowed requires further analysis.

[103] *See e.g., A Century of Nature: Twenty-One Discoveries that Changed Science and the World* xv-xvii (Laura Garwin and Tim Lincoln eds) available at https://press.uchicago.edu/Misc/Chicago/284158.html

[104] Matthew Sag, *Fairness and Fair Use in AI*, FORDHAM L. REV. (forthcoming) at https://papers.ssrn.com/sol3/papers.cfm?abstract_id=4654875

[105] *Id.*

[106] Campbell v. Acuff-Rose Music, Inc., 510 U.S. 569, 585 n. 18 (1994).

Even when one does not need permission to use a work, acts that can be characterized as theft to obtain access to the work do not do well in fair use analysis. In *Wright v. Warner Books*, the Ninth Circuit rejected an argument of bad faith and failure to gain permission in part because the material at issue had been given to the alleged infringer, "not stolen."[109] The court contrasted the situation, however, with the "sharp practices" the Supreme Court's condemned in *Harper & Row Publishers, Inc. v. Nation Enterprises*,[110] where the Nation used "*purloined*" letters instead of biding on the market for the material.[111] Furthermore, the Google Books line of cases which support using copyrighted material for software training and non-expressive uses noted, "Google provides the libraries with the technological means to make digital copies of books that *they already own*."[112]

---

[107] *Id.*; *accord* Castle Rock Entertainment, Inc. v. Carol Pub. Group, Inc., 150 F.3d 132, 146 (1998); Wright v. Warner Books, Inc., 953 F.2d 731, 737 (1991) ("lack of permission is 'beside the point'")
[108] As discussed infra, the world where copyright holders dictate uses for software is not desired as it would create unending problems for software development and use in the future. *See infra* Part III.A.3.
[109] Wright v. Warner Books, Inc., 953 F.2d 731, 737 (1991) ("lack of permission is 'beside the point'")
[110] 471 U.S. 539 (1985)
[111] Harper & Row Publishers, Inc. v. Nation Enterprises, 471 U.S. 539, 563 (1985)
[112] Authors Guild, Inc. v. Google Inc., 954 F.Supp.2d 282, 203 (2013)

Legal scholars point to touchstone cases about reverse engineering as supporting the ways LLM companies have gathered data, but those cases are different than the book corpora at issue.[113] In *Sega v. Sega Enterprises Ltd. v. Accolade, Inc.*, the defendant reversed engineered video games by using "commercially available copies" of the plaintiff's work; not stolen.[114] In *Sony Computer Ent. v. Connectix Corp.*, the defendant "purchased a Sony PlayStation console and extracted the Sony BIOS from a chip inside the console."[115] Thus, cases about reverse engineering software did not involve theft as the reverse engineer could buy the material as step one in the reverse engineering process. The court in *A V Ex Rel. Vanderhye v. iParadigms, LLC*, held the use of essays to train plagiarism software was fair use.[116] But again, the access to the underlying copyrighted material was lawful.[117]

Although you may not need permission to use copyrighted material when the use is fair use, cases that support that position involved instances where access was legally obtained.[118] Even when a court has found fair use when access was illegal, the fact of illegal access weighed *against* fair use. In *NXIVM Corp. v. Ross Inst.,* the court found that posting a small portion of a single manuscript to the Web, was fair use because the use was transformative, but still noted that when "access to the [work at issue] was unauthorized or was derived from a violation of law or breach of duty, this consideration weighs in favor of [the] *plaintiffs*" and against fair use.[119] Although the case seems to allow unauthorized access as long as the use is transformative and overall fair use, the access and use was on a small scale as compared to the book corpora used by OpenAI. A rule allowing access to books, even unauthorized copies of the books, aids AI research and fits within fair use ideals; yet the assumption that OpenAI was proper in following academic customs has problems.

[113] *See e.g.,* Pamela Samuelson, Christopher Sprigman, and Matthew Sag, *The FTC's Misguided Comments on Copyright Office Generative AI Questions*, PatentlyO, December 11, 2023, at https://patentlyo.com/patent/2023/12/misguided-copyright-generative.html; *but see* Pamela Samuelson, *Generative AI meets copyright* 381 *Science* 381, 158, 158 (2023) (noting propriety of using data from the "open Internet" which suggest lawful access).
[114] 977 F.2d 1510, 1514 (9th Cir. 1992).
[115] 203 F.3d 596, 601 (9th Cir. 2000).
[116] 562 F. 3d 630 - Court of Appeals, 4th Circuit 2009
[117] AV v. iParadigms, Ltd. Liability Co., 544 F. Supp. 2d 473, 478-480 (E. Dist. VA 2008) (describing the nature of the gathering the data and the contract authorizing use of the data). The software to detect plagiarism was built on material where the student consented to the use. One can argue that the consent was coerced or unconscionable, but the court looked at the access as legitimate.
[118] Castle Rock Entertainment v. Carol Pub. Group, Inc., 955 F.Supp. 260, 262 (1997) (detailing ways defendant built trivia list for the T.V. show, *Seinfeld*, including watching broadcasts and videotape recordings). When 2 Live Crew parodied Roy Orbison all they had to do was listen to the song and perhaps buy the sheet music; both of which are legal access to the work.

OpenAI's use of a little more than 290,000 titles from "shadow web libraries" that host illegal copies of copyrighted books[120] is at least bad optics, and at worst it's bad management.[121] Unlike most academic researchers, OpenAI had vast amounts of money for its research but chose not to spend the money on its data. In contrast, in the Google Books case, Google worked with libraries that had purchased the books. Google also invested in copying technologies to digitize the books.[122] By one estimate, Google spent $400 million to digitize 25 million books or roughly $16 per book.[123] Rather than taking a shortcut, OpenAI could have paid roughly $20 per new hardcopy of each book for a total of $5,800,000.[124] OpenAI would then have had to invest in scanning technology and hire people to operate the scanners. As folks at the venture capital firm Andreessen Horowitz put it, the cost of hardware and software to build and train models is high but "within reach of a well-funded start-up"; the same logic should apply to the data that goes into the model.[125] It is expensive work, but expenses are part of the marketplace.

---

[119] NXIVM Corp. v. Ross Inst., 364 F.3d 471, 478, 480 (2d Cir. 2004).
[120] https://qz.com/shadow-libraries-are-at-the-heart-of-the-mounting-cop-1850621671
[121] *Cf.* James Grimmelmann, *Copyright for Literate Robots*, 101 IOWA L. REV. 657, 674-675 (2016) (noting an approach that "ignores robots" encourages copying to feed the robots including in an "underground economy")
[122] James Somers, *Torching the Modern-Day Library of Alexandria*, THE ATLANTIC, April 20, 2017, at https://www.theatlantic.com/technology/archive/2017/04/the-tragedy-of-google-books/523320/
[123] *Id.*
[124] *See* SLJ's Average Book Prices (showing range of new book prices for libraries between 2021 and 2022) at https://s3.us-east-1.amazonaws.com/WebVault/SLJ/EDIT22/PDF/Avg-Book-Prices-Full-Chart-2022.pdf. By extension, after several lawsuits were filed, OpenAI began negotiating with news publishers about payments for access to their material. *See* Sharin Ghaffary, OpenAI's COO Pitches Startup As Friendly Partner to Publishers, Bloomberg, January 11, 2024, (noting closed deals with Associated Press and Alex Springer and ongoing negotiations with other publishers) at https://www.bloomberg.com/news/newsletters/2024-01-11/openai-s-coo-pitches-startup-as-friendly-partner-to-publishers

The *Authors Guild* case provides clear guidance that you can copy all of a given piece of copyrighted material for training; it does not bless all the possible ways you might obtain that material. OpenAI's attempt to cloak itself within what is allowed in academic settings by creating a private academic-styled place with non-profit ideals is no shield once the company took a commercial turn. The simplest lesson for LLM companies is to make sure to know from where your data comes and that it is a legally sanctioned source.

Even if fair use allows academic and, possibly non-academic, researchers to use book corpora available online, arguably even unauthorized ones, researchers should pay attention to how models built on such resources will be used. If the research is published and used in limited academic settings, the work should be protected under fair use. If the researcher has visions of starting a small company that might offer commercial products or be being bought, as when Google bought Deep Mind, things change.

Non-academic research must understand that it cannot assume that basic research is viable for commercial results. Even if you give credence to OpenAI's initial claims that it was a non-profit pursuing AGI for non-commercial outcomes, and only along the way saw those outcomes had potential to be commercial services, that is no defense to offering a system that used memorization and illicit data for its products. This point applies to Google, Microsoft, and any number of companies that have excellent, formerly academic researchers and forget that they are no longer in an academic setting.

---

[125] Guido Appenzeller, Matt Bornstein, and Martin Casado, *Navigating the High Cost of AI Compute*, ANDREESSEN HOROWITZ, April 27, 2023

When research is put into practice as a product, the inputs and the outputs matter. If those inputs or outputs violate intellectual property law, there can be repercussions. Questions about what role the inputs played in building the software will arise and fines and injunctions on using the software if it continues to yield infringing outputs are possible. In addition, as the legal landscape evolves software makers may find there is a sort of lien or shadow on their product where anyone with a copyright on the underlying data might argue the model is unauthorized and so payment is owed.

Furthermore, OpenAI's actions highlight a problem for academic researchers' legal/ethical dilemma with using books for AI research: access to copyrighted material is going to be either expensive or come from illegal copies online.[126] Even if Professor Sag is correct that as a practical matter requiring lawful access to copyrighted material such as books is "churlish," which we take to mean copyright holders will either say no, attach undesired limits, or ask exorbitant amounts, that doesn't solve whether using illegal copies of books is fair use and by extension, how researchers can continue to use shadow libraries without fear of lawsuits.[127] It may be that there will be clearer law that using shadow libraries is fair use. We think, however, that a better option is possible: a centralized common access point for copyrighted books. Before we explain that solution, we turn to another important input for LLMs, the Internet.

---

[126] *Cf.* Sag, *supra* note 104 (noting possibility no-one was willing to sell access to their books and so turning to shadow libraires is an understandable option).
[127] Professor Sag asserts "prohibiting academic research on illegal text corpuses will generally not benefit copyright owners or further the interests copyright is designed to promote." *Id*. Although we tend to agree with his point, the situation is not resolved.

2. The Internet as LLM Inputs

Data gathering may include book corpora, but it will also include even larger amounts of copyrighted work: writing and images on the web. As is well known, these sources are also covered by the overbreadth of copyright protection.[128] That data is vital because of its size. For example, in one training of GPT-3, 432 billion tokens out of about 499 billion total tokens came from the web.[129] Before OpenAI's launch of ChatGPT and DALL-E, almost anyone on the Web wanted their Web pages indexed and findable, so they allowed their sites to be crawled by search engines.[130] That crawling is what also allows someone to scrape the material for software training. But what if a website does not wish to be crawled? By custom, a website can place "robots.txt" into its web pages as a signal to search engines and other crawlers not to crawl the site. Obeying robots.txt instructions is, however, voluntary.[131] Indeed, some archive groups ignore the signal because it defeats their preservation projects.[132] Yet honoring requests not to crawl can provide a large dataset, for now.

[128] See e.g., Henderson, et al., *supra* note 1("Under United States ("U.S.") law, copyright for a piece of creative work is assigned "the moment it is created and fixed in a tangible form that it is perceptible either directly or with the aid of a machine or device" (U.S. Copyright Office, 2022). The breadth of copyright protection means that most of the data that is used for training the current generation of foundation models is copyrighted material.").
[129] Tom Brown, et al., *Language Models Are Few-shot Learners*. 33 ADVANCES IN NEURAL INFORMATION PROCESSING SYSTEMS 1877 (2020).
[130] *See e.g.,* Rebecca Bellan, *News Outlets Are Accusing Perplexity of Plagiarism and Unethical Web Scraping*, TECHCRUNCH, July 2, 2024 (explaining web scraping as automated crawlers often used by search engines so that web sites can be "included in search results"); *Cf.* Pamela Samuelson, *Generative AI Meets Copyrigh*" 381 SCIENCE 158, 159 (2023)
(arguing use of "a dataset consisting of 5.85 billion hyperlinks that pair images and text descriptions from the open internet" likely lawful given the dataset was made by a European non-profit and protected by EU law on using copyrighted material for text and data mining).
[131] As discussed below, whether ignoring robots.txt or other signals violates the law is an evolving question. The point here is that the custom has relied on voluntary compliance for the most part. *See e.g.*, Kali Hays, *A New Web Crawler Launched by Meta Last Month Is Quietly Scraping the Internet for AI Training Data*, FORTUNE, August 20, 2024 (noting obeying robots.txt is "not enforceable or legally binding in any way") at https://finance.yahoo.com/news/crawler-launched-meta-last-month-225937771.html

If you want a dataset that represents a wide range of voices and perspectives, an often-used dataset, CommonCrawl, is a strong candidate to be in your training dataset; but that resource is under attack. CommonCrawl claims to obey robots.txt and nofollow signals.[133] Its November December 2023 crawl release still has "3.35 billion web pages (or 454 TiB of uncompressed content)."[134] Thus, it seems researchers can access a robust dataset via CommonCrawl and respect online customs regarding how to use data. That situation is changing.

Some important web sites are adding a new "do not train" signal to their web pages[135] and in some cases their terms of service.[136] That is, web sites are indicating that people can scrape and index the site but not use that data to train software--what could be called a "scrape but don't train" signal. That change raises new questions for academic and corporate research. If CommonCrawl offers a dataset that has "do not train" data, will users know that? Is CommonCrawl liable for offering such a dataset even though CommonCrawl can't police how the data will be used? What, if anything, can a user know about the rules around CommonCrawl data going forward? These questions and more will plague AI research if restrictive "do not train" terms of service persist and spread.

---

[132] Brad Jones, *Internet Archive Will Ignore Robots.txt Files to Keep Historical Record Accurate*, DIGITAL TRENDS, (April 24, 2017) at https://www.digitaltrends.com/computing/internet-archive-robots-txt/#ixzz4gQYOqpUi
[133] https://commoncrawl.org/faq
[134] https://commoncrawl.org/blog/november-december-2023-crawl-archive-now-available
[135] Ariel Bogle, *New York Times, CNN and Australia's ABC Block OpenAI's GPTBot Web Crawler from Accessing Content*, THE GUARDIAN, (August 24, 2023) https://www.theguardian.com/technology/2023/aug/25/new-york-times-cnn-and-abc-block-openais-gptbot-web-crawler-from-scraping-content
[136] The NY Times Terms of Service Section 4.1(3) Prohibited Uses of the Service states "Without NYT's prior written consent, you shall not: use the Content for the development of any software program, including, but not limited to, training a machine learning or artificial intelligence (AI) system" https://help.nytimes.com/hc/en-us/articles/115014893428-Terms-of-Service

The move to prevent most, if not all, ability to scrape Web data to "train" software is a mistake. Publishers and authors have understandable concerns over whether their material is being used to generate works that compete with their human creative *outputs*. Outputs are not inputs. Yet, rather than focus on outputs, the response to OpenAI's products has been to limit access to inputs. The NY Times, CNN, Reuters, and other major news organizations are blocking at least OpenAI's scraping tool[137] with the NY Times blocking Common Crawl as well. The BBC is exploring how to use generative AI systems in its reporting; and it is blocking LLM companies from crawling BBC sites to fuel a company's software development.[138] The BBC's position is consistent with the emerging practice of large entities using their material to build an LLM internally.[139] The choice is ironic. The technology behind using your own data to have an internal LLM relies on research that needed access to the sites now trying to cut off that access. In addition, in the age of data journalism the broad bans on access to site data are either short-sighted or hypocritical.

The definition of prohibited training can be so broad as to be absurd. Training is such a fundamental term that it includes many ways almost any machine learning model would work. The NY Times, a leader in data journalism, offers a good example of over-reaching terms of service. Suppose you wanted to study the NY Times to detect bias in reporting. You may think the paper is too liberal or racist or favors a certain political view or disfavors a view in subtle ways.[140]

---

[137] *See supra* Bogle, note 135.
[138] Emilia David, *The BBC Is Blocking OpenAI Data Scraping But Is Open to AI-powered Journalism*, THE VERGE, (October 6, 2023) https://www.theverge.com/2023/10/6/23906645/bbc-generative-ai-news-openai
[139] For example, a large U.S. law firm used its material to create an internal Chatbot to aid in finding legal documents, drafting material, and knowledge discovery. *See* Bob Ambrogi, *Four Months After Launching Its 'Homegrown' GenAI Tool, Law Firm Gunderson Dettmer Reports On Results So Far, New Features, And A Surprise on Cost*, LAWSITES, (December 20, 2023) at https://www.lawnext.com/2023/12/four-months-after-launching-its-homegrown-genai-tool-law-firm-gunderson-dettmer-reports-on-results-so-far-new-features-and-a-surprise-on-cost.html
[140] James Bennet, *When the New York Times Lost Its Way*, The Economist, (December 14, 2023) at https://www.economist.com/1843/2023/12/14/when-the-new-york-times-lost-its-way

The language in reporting would be a way to examine and support your hypothesis. To test this hypothesis, you may want to train a classifier model or a language model just on NYT articles and perform statistical analyses. Under the terms of service, however, you would need the NY Times' written consent to access the site.[141] You would also need permission to use any automated software to gather data from the site.[142] Even if you had that data, you could not "use the Content for the development of any software program, including, but not limited to, training a machine learning or artificial intelligence (AI) system."[143] The words, *the development of any software*, encompasses writing script to analyze data. The Terms also prohibit caching and archiving the data.[144] Thus, if someone wanted to show how they arrived at their conclusions and that the analysis was valid, that person would fail. Without the data, a third party cannot verify results.[145] These restrictions beg the question, would the NY Times really obey its own Terms of Service? Perhaps not.

The NY Times has embraced data journalism and analysis especially through its UPSHOT section, and that work further shows why restrictive terms of service make little sense. The UPSHOT prides itself on analyzing data as part of its reporting. It describes itself as "Analysis that explains politics, policy and everyday life, with an emphasis on data and charts."[146] Part of its mission is "unearthing data sets—and analysing [sic] existing ones—in ways that illuminate and, yes, explain the news.[147] One of the Section's Top 10 stories of 2023 was Nate Cohn's piece, *6 Kinds of Republican Voters*.[148] As he said, "the first output of [the] statistical analysis — called a cluster analysis — ... didn't make sense to me [], at first. It took a while to make sense of how my algorithm was segmenting Republicans."[149] In the article, one group, Libertarian Conservatives, was perplexing because it was "near the middle of the pack on almost every set of issues."[150] Yet his algorithm "set them apart."[151] The words "first output" and "cluster analysis" are about algorithms trained on data.[152] Indeed, Cohn and his team had to analyze the algorithm and perhaps rerun the data to generate new outputs beyond the first output to make sense of the data and what it seemed to show. That work *is* software development and training. Cohn used a survey and so developed his data in-house, but what if a good data journalist gathered data to analyze it? That's what the Times did with another piece.

---

[141] NY Times Terms of Service Section 4.1(1) at https://help.nytimes.com/hc/en-us/articles/115014893428-Terms-of-Service
[142] NY Times Terms of Service Section 4.1(2) at https://help.nytimes.com/hc/en-us/articles/115014893428-Terms-of-Service
[143] NY Times Terms of Service Section 4.1(3) at https://help.nytimes.com/hc/en-us/articles/115014893428-Terms-of-Service
[144] NY Times Terms of Service Section 4.1(5) at https://help.nytimes.com/hc/en-us/articles/115014893428-Terms-of-Service
[145] As discussed below, a data repository may allow testing, but that presumes there is such a repository. If you are scraping data and/or assembling data from one or several sources, you are the likely repository of the data and so need a copy of the data.
[146] https://www.nytimes.com/newsletters/upshot; Natalie Gill, *New York Times Launches Data Journalism Site The Upshot*, THE GUARDIAN, (April 22, 2014) ˆ https://www.theguardian.com/media/2014/apr/22/new-york-times-launches-data-journalism-site-the-upshot

The UPSHOT used external data and analyzed it for an article examining "How Formulaic" Hallmark and Lifetime movies are, "Based on data from IMDb, internet recaps and our own viewing."[153] IMDB's Terms of Service include language requiring written permission for any commercial purpose.[154] The reporters used IMDb's API to access the Metadata. Whether the NY

---

[147] *See* Gill, *supra* note 147
[148] The Upshot Staff, *10 Data Points and Documents That Made US [ponder emoji] in 2023*, NY TIMES, (December 30, 2023) at https://www.nytimes.com/interactive/2023/12/30/upshot/2023-year-in-review.html
[149] *Id.*
[150] Nate Cohn, *The 6 Kinds of Republican Voters*, NY TIMES, (August 17, 2023), at https://www.nytimes.com/interactive/2023/08/17/upshot/six-kinds-of-republican-voters.html
[151] *Id.*
[152] *Cf.* Vladimir Estivill-Castro, *Why So Many Clustering Algorithms: A Position Paper*, 4 ACM SIGKDD Explorations Newsletter 65 (2002) ("Clustering is a central task for which many algorithms have been proposed").
[153] Alicia Parlapiano, *Just How Formulaic Are Hallmark and Lifetime Holiday Movies? We (Over)analyzed 424 of Them*, NY TIMES, (December 23, 2023) at https://www.nytimes.com/interactive/2023/12/23/upshot/hallmark-lifetime-christmas.html
[154] https://www.imdb.com/conditions ("The IMDb Services or any portion of such services may not be reproduced, duplicated, copied, sold, resold, visited, or otherwise exploited for any commercial purpose without express written consent of IMDb.")

Times paid the $150,000 plus metered costs for basic API access or used the 1-month free trial is unclear.[155] Nonetheless, the NY Times seems to have had permission to gather the data and analyze it. But "Internet recaps" likely come from a range of websites that may not be happy to have their content scraped or gathered in some other way. Whether the reporters checked those sites' Terms of Service statements is unknown. Even if the sites did not have Terms of Service statements, the ethos of not using another site's data for your own purposes would seem to indicate that the reporters ought not to have used those recaps or perhaps asked permission as a courtesy. Even if all the data was gathered with permission, did the reporters let sources know they would use software to analyze the data? What if the reporters wrote a little bit of code to do that? As with Mr. Cohn's work, that is "development of software." The code might be useful for future stories and generating new insights, which is yet another way the work is development.

In general, permission-driven systems for Internet available material raise many problematic questions. Will the NY Times, other news agencies, NGOs, and other civic groups have to get permission for all their data gathering in reporting or any attempt to use the Internet to study society? Will they need permission to build software to analyze that data? Will that permission allow further use of the data and software, or will the permission be limited and short-term? What about vital advances in AI related to vision research or cybersecurity? Will they be encumbered if they used the Internet as part of their results? What about academic research? Will academics need to clear all material for research in language, vision, robotics, and other AI research? Will that research also have limits on further use? These questions are maddening and a

---

155 https://aws.amazon.com/marketplace/search/results?FULFILLMENT_OPTION_TYPE=DATA_EXCHANGE&CREATOR=0af153a3-339f-48c2-8b42-3b9fa26d3367&DATA_AVAILABLE_THROUGH=API_GATEWAY_APIS&filters=FULFILLMENT_OPTION_TYPE%2CCREATOR%2CDATA_AVAILABLE_THROUGH

clearance nightmare, but they follow from restrictive legal terms. Indeed, the history of commercial entities engaging in copyright licensing and clearances shows the problems of giving copyright holders too much power.

### 3. We Should Not Repeat Mistakes of Copyright Licensing History

A future where a copyright holder could argue that data was allowed for one type of training or software but not another would lead to a ball of confusion and litigation.[156] Consider the issues around music use in film and television. Licensing can be so tight that if a copyright holder allowed use of a song for TV broadcast, but not on a video format (VHS, DVD, Blu-ray, etc.) or for streaming, the show will be inaccessible.[157] For example, lack of clearance rights for music as used in visual media has stopped access to important work on civil rights and limited the ability of films and tv shows to be shown on streaming services.[158] Put simply, further expansion of copyright to cover use of copyrighted material for training would expand what, more than fifteen years ago, Michael Heller identified as a Gridlock Economy--one where property owners act as Robber Barons over-charging for access to the commons.[159]

---

[156] *Cf*. Tasini v. New York Times, 206 F.3d 161, 135 (2d. Cir. 2000) (holding that a copyright license is specific and a license that did not include "electronic databases" was violated when publishers put the material into such databases).

[157] *Cf*. Tasini v. New York Times, 206 F.3d 161, 135 (2d. Cir. 2000) (holding that a copyright license is specific and a license that did not include "electronic databases" was violated when publishers put the material into such databases).

[158] Josef Adalian*, Why Are These Classic Shows Nowhere to Be Found on Streaming?*, Vulture (November 18, 2018) at https://www.vulture.com/2016/11/why-cant-these-shows-be-found-on-streaming.html ("And then there's what's often the biggest obstacle to getting a show on to a streaming platform: music rights.")

[159] *See* HELLER, *supra* note 15.

Expanding copyright to include software training would also ignore what Jerome Reichmann and Ruth Okediji figured out more than ten years ago. Copyright laws already pose a threat to scientific research, especially research that uses digital techniques such as "data-mining techniques, and other automated knowledge discovery tools."[160] Allowing the copyright industry to control inputs to and dictate terms of use for software development would make the industry gatekeepers of future uses of software and offer too much power to an industry that already overreaches with over-long copyright terms and extraordinary barriers to access to knowledge.

Furthermore, the shift by some LLM companies to licensing rights to train their software,[161] reveals problems with deferring to the copyright industry. The move does not solve whether someone should be required to do so. Indeed, fair use law maintains that licenses are not required when the use is fair use;[162] yet some groups are arguing because one could license, one must.[163] In addition, given the recent concerns over antitrust and technology in general, deference to copyright power and industry as the FTC's recent statements about AI training and copyright is a mistake.[164]

The FTC's myopia about technology companies simply misses the abuse and anti-competitive nature of the copyright industry. As part of that myopia, the FTC raised fears about

---

[160] Jerome H. Reichman and Ruth L. Okediji, *When Copyright Law and Science Collide: Empowering Digitally Integrated Research Methods on a Global Scale*. 96 MINN. L. REV. 1362, 1368 (2012).
[161] *See e.g.*, Michael Nunez, *OpenAI Strikes Content Deal with Condé Nast, Raising Questions About Future of Publishing*, VENTURE BEAT, August 20, 2024 at https://venturebeat.com/ai/openai-strikes-content-deal-conde-nast-future-of-publishing/; Bellan, *supra* note 65; Benj Edwards and Ashley Belanger, *Journalists "deeply troubled" by OpenAI's content deals with Vox, The Atlantic*, ARS TECHNICA, May 31, 2024 at https://arstechnica.com/information-technology/2024/05/openai-content-deals-with-vox-and-the-atlantic-spark-criticism-from-journalists/
[162] *See e.g.*, Damon Beres, *A Devil's Bargain With OpenAI*, THE ATLANTIC, May 29, 2024 (quoting Anna Bross of the Atlantic that the license is not for "syndication" but and that display of content must still be within fair use) at https://www.theatlantic.com/technology/archive/2024/05/a-devils-bargain-with-openai/678537/
[163] *See e.g.*, Cho, *supra* note 92.
[164] COMMENT OF THE UNITED STATES FEDERAL TRADE COMMISSION, Artificial Intelligence and Copyright, Docket No. 2023-6, October 30, 2023 available at https://www.regulations.gov/comment/COLC-2023-0006-8630

"training an AI tool on protected expression without the creator's consent."[165] The statement assumes author's and artists will be compensated, but the data is concentrated in large publishers and many actual creators are skeptical of the current licensing deals both because of where the money may go and for fear that the deals with foster software that might reduce the need for the writers who created the material in the first place.[166] In addition, the handwave at taking care of creators ignores the reality of the market. If one moves to requires licensing, the amount of data needed means the costs become millions if not billions of dollars.[167] As one researcher noted, those "governing content" have strong reasons "to lock up their materials." As that happens early movers who grabbed data are "bless[ed]" and the shift "pull[s] up the ladder so nobody else can get access to data to catch up."[168] In short, assuming and/or agreeing that all copyrighted data needs a license to train models fosters the exact thing the FTC claims not to want: a limited and anti-competitive market that also probably does not pay authors. Last, the view misses the issues around fair use which arguably is designed to thwart the anti-competitive power of copyright. Well-funded companies can afford to experiment with a mix of licensing and fair use arguments/litigation. Other actors cannot.

---

[165] COMMENT OF THE UNITED STATES FEDERAL TRADE COMMISSION, Artificial Intelligence and Copyright, Docket No. 2023-6, October 30, 2023 available at https://www.regulations.gov/comment/COLC-2023-0006-8630

[166] *See e.g.*, Damon Beres, *A Devil's Bargain With OpenAI*, THE ATLANTIC, May 29, 2024 (noting the issues around AI as a threat to writers and the way previous deals between technology companies and journalism ended up not benefitting the news outlet) at https://www.theatlantic.com/technology/archive/2024/05/a-devils-bargain-with-openai/678537/; Nunez, *supra* note 161; Benj Edwards and Ashley Belanger, *Journalists "Deeply Troubled" by OpenAI's Content Deals with Vox, The Atlantic*, ARS TECHNICA, May 31, 2024 at https://arstechnica.com/information-technology/2024/05/openai-content-deals-with-vox-and-the-atlantic-spark-criticism-from-journalists/

[167] Kyle Wiggers, *AI Training Has a Price Tag That Only Big Tech Can Afford*; TechCrunch, June 1, 2024 (noting the growing need for more data to train LLM models and the multi-billion dollar market emerging to meet that demand)

[168] *Id.*

Even if fair use doctrine protects academic research and non-expressive uses of copyrighted material, a custom of do not crawl code and restrictive terms of service creates moral and ethical mayhem for legitimate uses. Sites are increasingly signaling that they don't wish to be crawled and/or used for study and development of AI. Thus, researchers will be forced to ignore custom and perhaps transgress online ethical standards.

Evolving legal standards pose problems too. Some caselaw suggests failing to use metatags such as "no archive" creates an implied license.[169] By extension, if you use the "no-archive" metatag, you have denied the license. At least one court has held that implied licenses are particular; a license to use web content for one purpose *does not* give license for another purpose.[170] Worse, potential legal action can stop research and innovation before they start.[171] Research institutions may require clearance for all usage simply as a risk-averse, fear of litigation approach.[172]

The focus on a sliver of LLM outputs recklessly offered by industry is myopic and misses the broader ways LLMs offer advances in AI research and business productivity. Paragraphs, book chapters, essays, and images that may violate copyright are a subset of potential outputs from LLMs. Search, spell-checking, auto-complete, customer support systems, analysis of legal and financial documents, language translation software,[173] are but a sample of the sorts of systems that leverage LLMs. The sort of fundamental breakthroughs in NLP, neural networks, self-improvement, randomness and creativity, computer vision, and other AI research areas also need access to the copyrighted material on the web.[174] In short, more data and high-quality data are important for future advances and future applications of LLMs as well as AI research in general.

[169] Field v. Google Inc., 412 F. Supp. 2d 1106, 1115–17 (D. Nev. 2006); *cf.*, Samuelson, *supra* note 5 (relying on access to Open Internet as part of the argument for fair use when training on Internet images).
[170] MidlevelU, Inc. v. ACI Info. Grp., 989 F.3d 1205, 1217-1218 (11th Cir. 2021).
[171] One can think of the issue as a chilling effect. The mere threat of lawsuits and high damages has a history of preventing action especially when intellectual property is at stake. *See e.g.*, Deven R. Desai, *Speech, Citizenry, and the Market: A Corporate Public Figure Doctrine*. 98 MINN. L. REV. 455, 477-478 (2013) (connecting chilling effects of tort lawsuits in *Times v. Sullivan* to trademark enforcement suits)
[172] *Cf.* Cambridge University Press v. Becker, 863 F. Supp. 2d 1190 - Dist. Court, ND Georgia 2012) (publishers sued University System of Georgia and each Regent as responsible for the university's copyright policies).

[173] *See e.g.,* CellStrat, *Real-World Use Cases for Large Language Models (LLMs)*, MEDIUM, (April 25, 2023) https://cellstrat.medium.com/real-world-use-cases-for-large-language-models-llms-d71c3a577bf2

Ex ante restrictions on access to copyrighted material for non-expressive, non-infringing purposes will recreate problems that plague use of copyrighted material. X, formerly Twitter, already offers an approach of rather useless or high-cost API access. The $100 per month access provides only "0.3 percent of [the data researchers] previously had free access to in a single day."[175] The enterprise access plan is apparently $42,000 per month and still does not allow enough access for some major studies and analytical tools.[176] X's approach shows the way access can be cut off and/or over-priced for just one, centralized platform. Like X, Getty Images, the NY Times, Penguin Random House, HarperCollins, Simon & Schuster, Hachette, etc. could all demand payment for any access.[177] Denying or charging exorbitant fees is not the only problem with ex ante access. As with the commercial context, once the copyright industry can exert power over access to copyrighted material for training purposes, it will likely add terms regarding what training is allowed as has already happened with the NY Times' terms of service.

---

[174] *See* McCarthy et al., *supra* note 20.
[175] Justine Calma, *Twitter Just Closed the Book on Academic Research*, THE VERGE, (May 31, 2023) at https://www.theverge.com/2023/5/31/23739084/twitter-elon-musk-api-policy-chilling-academic-research
[176] *Id*. ("even its "outrageously expensive" enterprise tier, the coalition argued, wasn't enough to conduct some ambitious studies or maintain important tools")
[177] *See e.g.*, Cho, *supra* note 92.

## B. Data Outputs – Simpler Fixes

Past clashes between platform companies and the copyright industry show a simpler path forward. Commercial companies must be more careful. Had OpenAI followed the customs and legal guidelines about copyrighted material being placed online, the debate around copyrighted material as inputs might not have been resurrected. As we have said before, outputs matter. Indeed, as two computer scientists offer "it's a bit surprising that OpenAI has let [issues around preventing its outputs from offering from copyright-violating results] get this far."[178] AI companies must test software to see whether outputs offer outright IP infringement. AI companies must also use common techniques for filtering to mitigate potential harms. Neither of these practices means that a company can identify all the possible negative outcomes.[179] But relying on ex post fixes and generosity to identify harms such as paywall bypassing suggests a company that is either clueless about classic issues between platform and content industry or doesn't care.[180] The question is "Did a company at least try to mitigate harm?"[181]

A consistent rule in intellectual property rights versus platform companies is that platform companies should work to mitigate intellectual property harms. The Google Books case again offers instructions. Unlike OpenAI's approach which yielded a full page or pages of a book,

---

[178] Arvind Narayan and Sayash Kapoor, *Generative AI's End-run Around Copyright Won't Be Resolved by the Courts* AI SNAKE OIL at https://www.aisnakeoil.com/p/generative-ais-end-run-around-copyright
[179] *Cf.* Desai and Kroll, *Trust But Verify*, (explaining the limits of detecting all possible results from software)
[180] *See* Narayan and Kapoor, *supra* note 178 (noting the speed with which OpenAI fixed a ChatGPT web browsing feature that could "bypass paywalls" once the computer scientists mentioned it on X)
[181] Katherine Lee, A. Feder Cooper, and James Grimmelmann, *Talkin'Bout AI Generation: Copyright and the Generative-AI Supply Chain* (2003). *arXiv preprint arXiv:2309.08133*. ("precedents have come to set expectations— among copyright owners, in the technology industry, in the copyright bar, and in the judiciary — for what legally "responsible" behavior by an online intermediary looks like. A generative-AI service operator that does not appear to be making a good-faith effort to achieve something like this system may strike a court as intending to induce infringement").

Google's Book Project used snippets of books and has done so for almost 20 years.[182] Snippets are about three lines of text and Google placed guardrails on results so users could not prompt the system to reveal so many snippets that a user could see whole pages.[183] Instead of throwing items out the lab window onto the Web to see what would happen, OpenAI should have tested its software in-house and asked questions about what it might do. Indeed, any company should not only test its software but also ask, what this software is good for or supposed to do?[184]

Recent Internet technology research and product history illustrates what to do and what not to do. When Google built Street View, its researchers sucked up wifi data including user emails, passwords, and other information.[185] Some of that data was not needed for the product but was available and so scooped up.[186] Google's initial position was that things visible in public, including data that unsuspecting people did not encrypt, are by definition not private, and so could be gathered as part of the project.[187] After a multi-state lawsuit, Google settled and admitted that its practices violated people's privacy.[188] Google is not alone in its research errors. Facebook researchers published an infamous paper about how they could create social contagion online.[189] The paper shows that newsfeeds can indeed affect how we feel. When there are fewer positive posts, people create more negative posts and fewer positive posts; when more negative posts are in someone's feed, the opposite occurs.[190] Understanding the dynamic means you could better govern social media, *but* it also shows that social media companies could manipulate users. In addition, by conducting the study on more than 689,000 users, the researchers transgressed an ethical boundary.[191] In contrast, as much as people want to decry Amazon's building an in-house hiring tool that disfavored hiring women, Amazon tested the software and chose *not* to use it because of its flaws.[192] Amazon's approach is the better one.

---

[182] Authors Guild v. Google 804 F.3d at 208-210 (detailing with approval the nature of snippet view).
[183] Id. at 210 (Google's program does not allow a searcher to increase the number of snippets revealed by repeated entry of the same search term or by entering searches from different computers. A searcher can view more than three snippets of a book by entering additional searches for different terms. However, Google makes permanently unavailable for snippet view one snippet on each page and one complete page out of every ten—a process Google calls "blacklisting.").
[184] *See* supra Part I.A.2 (discussing how academic benchmark goals differ from commercial ones).
[185] David Streitfeld, *Google Concedes That Drive-By Prying Violated Privacy,* NY TIMES (March 12, 2013) https://www.nytimes.com/2013/03/13/technology/google-pays-fine-over-street-view-privacy-breach.html
[186] *Id.*
[187] *Id.*
[188] *Id.*
[189] Adam Kramer, Jamie E. Guillory, and Jeffrey T. Hancock, *Experimental Evidence of Massive-scale Emotional Contagion Through Social Networks*, 111 PROCEEDINGS OF THE NATIONAL ACADEMY OF SCIENCES OF THE UNITED STATES OF AMERICA 8788 (2014).

Insofar as OpenAI wanted to offer a service that was quite accurate as a search tool, the possibility that its software might produce problematic results should not have surprised the company. Once the science went from building software to show general language prowess to providing accurate answers, which requires memorization, the outputs were more likely to infringe or be close to infringing. Prompts such as "write in the style of X," or "summarize Y book by Z author," have yielded results that are the basis of some of the current lawsuits.[193] In one case, rather than have limits on outputs, OpenAI allowed a user to ask for a summary of several chapters of an individual author's book because the system could be prompted to summarize Chapter 1, Chapter 2, and so on.[194] Although summaries may be legal, summarizing each chapter in detail runs into the third fair use factor, which assesses the amount or substantiality of the portion used, in spirit, if not the letter, of the law.

---

[190] *Id.*

[191] Charles Arthur, *Facebook Emotion Study Breached Ethical Guidelines, Researchers Say*, THE GUARDIAN (June 30, 2104) at https://www.theguardian.com/technology/2014/jun/30/facebook-emotion-study-breached-ethical-guidelines-researchers-say; Evan Selinger and Woodrow Hartzog, *Facebook's Emotional Contagion Study and the Ethical Problem of Co-opted Identity in Mediated Environments Where Users Lack Control*, 12 RESEARCH ETHICS 35 (2016).

[192] *Cf.* Desai, Gupta, Salem, *supra* note 99 (discussing how legal literature uses Amazon as an example of data bias and harm when in fact Amazon used good practices to manage its software).

[193] Wes Davis, *Sarah Silverman Is Suing OpenAI and Meta for Copyright Infringement*, THE VERGE (July 9, 2023) https://www.theverge.com/2023/7/9/23788741/sarah-silverman-openai-meta-chatgpt-llama-copyright-infringement-chatbots-artificial-intelligence-ai

[194] https://s3.documentcloud.org/documents/23869694/silverman-openai-complaint-exhibits.pdf

To be clear, LLMs are not like Napster or Grokster where entire songs were shared, but the way OpenAI released its software without better guardrails gives the impression of widespread copyright output problems. As two computer scientists put it, "[F]rom a practical perspective, the idea of people turning to chatbots to bypass paywalls seems highly implausible, especially considering that it often requires repeatedly prompting the bot to continue generating paragraph by paragraph. There are countless tools to bypass paywalls that are more straightforward."[195] Nonetheless, by not thinking about how prompts might cough up the entire text of a book like *Oh The Places You'll Go* or the first 3 pages of *Harry Potter and the Sorcerer's Stone*,[196] OpenAI was at best cavalier about its outputs and possibly reckless.[197]

Any company online must understand the issues around online copyright and should have known that users are likely to use any tool possible to gain access to copyrighted works, especially ones that claim to be quite concerned about the misuse of AI. That issue has been at the core of Internet law since the birth of the commercial Internet. Indeed, two laws, Section 230 of the Communications Decency Act (Section 230), and the Digital Millenium Copyright Act (DMCA) are well-known laws born of that tension and have been around since 1996 and 1998 respectively. Yet, the lessons learned from commercial practices since then evaded OpenAI's management.

The two acts have flaws but, in a way, they balance each other. Section 230 offers platforms, such as YouTube, protection for hosting third party content that could infringe IP rights. But the Digital Millenium Copyright Act requires platforms to take reasonable steps to delete or prevent access to infringing copyrighted material once the platform has notice from the copyright holder.[198] As a practical matter these two rules clash or at least could have looked absurd. For example, in *Viacom Int'l. v. YouTube, Inc.* the sheer number of potential infringements on You Tube meant the film, television, music and sports industries faced nearly 100,000 violations that they knew of.[199] The music industry sued, and part of the argument or ethos was that it could not be that Congress meant to offer a solution where, over time, a huge volume of unauthorized copies of music were on YouTube.[200] YouTube might have stuck to Section 230 protection which had been upheld in cases with similarly poor outcomes. Yet rather than rely solely on the law, YouTube built Content ID which aids music publishers in identifying potentially infringing material, deciding whether to send a notice and takedown letter, or choosing to monetize a use of copyrighted material that would otherwise be difficult to negotiate.[201] That investment meant Viacom limited their case to acts *prior* to ContentID's implementation.[202]

---

[195] *See* Narayan and Kapoor, *supra* note 178.
[196] *See* Henderson et al., *supra* note 1.
[197] *But see* Narayan and Kapoor, *supra* note 178, ("But from a practical perspective, the idea of people turning to chatbots to bypass paywalls seems highly implausible, especially considering that it often requires repeatedly prompting the bot to continue generating paragraph by paragraph. There are countless tools to bypass paywalls that are more straightforward.").

Online trademark rights offer another example of better management.[203] The online marketplace, eBay, hosts a huge number of sellers that are difficult to police and that sometimes sell counterfeit goods.[204] Tiffany sued eBay and argued that eBay knew of the counterfeits and benefitted from those sales.[205] That argument, however, failed because of eBay's proactive steps in building a system that let sellers provide eBay with evidence of counterfeit goods so that eBay could shut down such sellers. In addition, eBay spent close to $20 million a year on trust and safety which included having nearly 4,000 people on the trust and safety team, 70 of whom focused on counterfeiting issues.[206]

---

[198] 17 U.S.C. § 512
[199] 718 F.Supp.2d 514, 524 (2010)
[200] Viacom Intern., Inc. v. YouTube, Inc., 940 F.Supp.2d 110, 115 (S.D.N.Y. 2013) (rejecting Viacom's argument that the volume of clips on YouTube precluded safe harbor protection).
[201] Miguel Helft, *Judge Sides With Google in Viacom Video Suit*, NY TIMES, B1, June 24, 2010.

[202] Miguel Helft, *Judge Sides With Google in Viacom Video Suit*, NY TIMES, B1, June 24, 2010.

[203] *Accord*, *See* Henderson et al., *supra* note 1 (noting trademark law and online sales as an example of the law embracing mitigation strategies).
[204] Tiffany v eBay, 600 F.3d 93 (2010)
[205] Tiffany v eBay, 600 F.3d 93, 103 (2010)

Arguing that OpenAI is not hosting third party content and so safe harbors do not apply misses the point.[207] eBay was not protected by the safe harbors either; it simply chose to try to mitigate harm to start. The core lesson is the more a platform or service mitigates harm, the better.

Even with smarter business practices in place, issues around access to data will persist. Future access to copyrighted inputs will mean a shift in research practices and new approaches to balancing access to research inputs and the possible need to compensate some, but not all, copyright holders.

## IV. The Path Ahead

The reality is LLMs, such as ChatGPT, need not be a threat to copyright-based industries; yet the lawsuits and outcry that the copyright industry be paid persist.[208] Researchers and lawsuits have probed ChatGPT in aggressive ways that are properly characterized as attacks and found troublesome outputs. As the public found out about these edge cases, even the FTC took notice with some ill-advised statements about "generative AI."[209] Public scrutiny and outcry and the threat of lawsuits has forced OpenAI and by extension other so-called “generative AI” companies to build guardrails such as limiting the sorts of prompts a system will follow, fine tuning to prevent verbatim outputs, and output filtering.[210] Why then does the drumbeat for compensation and over-reaching statements from the FTC about “training an AI tool on protected expression without the creator’s consent” continue?[211] Part of the furor is wrapped up in issues far larger than copyright compensation such as the future of work, support for news reporting,[212] misinformation, etc.[213] But make no mistake, part of the argument from the copyright industry is its usual argument that all uses of copyrighted material must be compensated.[214]

---

[206] Tiffany v eBay, 600 F.3d 93, 99, 109 (2010)

[207] *See e.g.*, Lee et al., *supra* note 181 (offering detailed analysis about why Section 512 does not apply to generative AI systems).

[208] *See e.g.*, Cho, *supra* note 85.

[209] *See e.g.,* Samuelson, et al., *supra* note 113.

In essence the argument sounds in the Lockean approach to copyright where the labor behind creating copyrighted material merits treating the material the same way as real property.[215] That view further holds that any uncompensated use is free-riding, a logic that has been mainly rejected.[216] Nonetheless, whether access to illegal copies of books will be allowed as part of fair use analysis is an open question. Furthermore, as Pamela Samuelson, Christopher Jon Sprigman, and Matthew Sag note, failure to respect robots.txt or other "mechanisms" designed to stop scraping for training purposes might count against fair use.[217] Thus, exactly how researchers should proceed regarding Internet content is also unresolved. This section offers solutions for research access to books and then for internet content.

[210] *See* Narayan and Kapoor, *supra* note 178.
[211] COMMENT OF THE UNITED STATES FEDERAL TRADE COMMISSION, Artificial Intelligence and Copyright, Docket No. 2023-6, October 30, 2023 available at https://www.regulations.gov/comment/COLC-2023-0006-8630
[212] *See e.g.,* Narayan and Kapoor, *supra* note 178 (claiming the big issue around generative AI is ”the injustice of labor appropriation in generative AI” and offering that the way ChatGPT hides the NY Times as source material will erode traffic to news sites)
[213] COMMENT OF THE UNITED STATES FEDERAL TRADE COMMISSION, Artificial Intelligence and Copyright, Docket No. 2023-6, October 30, 2023 available at https://www.regulations.gov/comment/COLC-2023-0006-8630 (noting FTC concern over ”the risks associated with AI use, including violations of consumers’ privacy, automation of discrimination and bias, and turbocharging of deceptive practices, imposter schemes, and other types of scams”)
[214] *See e.g.*, Cho, *supra* note 92.
[215] *See e.g.,* Deven R. Desai, *The Life and Death of Copyright* 2011 WISCONSIN L. REV. 219, 245 (2011) (tracing Lockean labor theory ideas in copyright law). There are other theoretical foundations for property. *See e.g.* Margaret Jane Radin, Property and Personhood, 34 STAN. L. REV. 957 (1982)). But as Professor Madhavi Sunder offers, “Assertions of power over one’s own identity necessarily lead to assertions of property ownership. . . . Property enables us to have control over our external surroundings. Seen in this light, it is not enough to see all claims for more property simply as intrusions into the public domain and violations of free speech. Instead, we may begin to see them as assertions of personhood.” Madhavi Sunder, *Property in Personhood*, in RETHINKING COMMODIFICATION: CASES AND READINGS IN LAW AND CULTURE 164, 170 (Martha M. Ertman & Joan C. Williams eds., 2005). Thus the theoretical foundation does not change the property instinct and the demand for control over the property.

## A. Maintaining and Expanding Access to Books

Large book corpora are vital to LLM research, but access to copies of books for legitimate software training is scare. You can reject the idea that all uses and access to books must be with permission and still recognize that using "shadow libraries" with two to three hundred thousand books raises problems. Such illicit access may not be fair use.[218] As matter of ethics and the optics of fairness, the outcry and uncertainty over the use of shadow libraries is loud regardless of whether the researchers are academics or non-academic. The issue needs to be resolved, and we believe there are a few possible solutions for using book corpora going forward.

[216] *Id.*; *see also*, Mark A. Lemley, *Property, Intellectual Property, and Free Riding*, 83 TEX. L. REV. 1031, 1032 (2005) (arguing that use of "the rhetoric of real property, with its condemnation of .'free riding.' By those who imitate or compete with intellectual property owners." has resulted in "a legal regime for intellectual property . . . in which courts seek out and punish virtually any use of an intellectual property right by another.")
[217] Pamela Samuelson, Christopher Jon Sprigman, and Matthew Sag, COMMENTS IN RESPONSE TO THE COPYRIGHT OFFICE'S NOTICE OF INQUIRY ON ARTIFICIAL INTELLIGENCE AND COPYRIGHT at 26 (October 30, 2023).
[218] *See supra* notes 109 to 117 and accompanying text.

Fair use presents possibilities but is unlikely to solve all the issues. Courts might recognize using such corpora is fair use for all software training.[219] This approach would track the idea that you need not ask permission to use copyrighted material when the use is fair. That approach would allow a broad rule that avoids the myriad issues that will inevitably follow if the copyright industry is able to control licensing and restrictions on types of use. It would also leave open the issue of whether the outputs are infringing, in which case fair use does not automatically apply. That outcome would balance the issue, because the copyright industry could vindicate rights when outputs that violate copyright law are created. But as we have explained, the scale and nature of access via shadow libraries means this outcome is not obvious or guaranteed.

Courts may hold that you must have legitimate access to the underlying work even when the use is fair. In this approach, you do not have to ask permission, but as with 2 Live Crew and other cases where permission was not required, you would have to have legal access to the material.[220] If you wanted to use digital books, which might carry terms of service about licensing and restrictions on use, courts should allow someone to buy the copy and then ignore such terms. As such, courts would have to be clear that evading digital copy controls, including using tools to circumvent Digital Rights Management controls, is allowed for such training as part of fair use doctrine. Researchers could instead buy and scan physical copies of the books. That would be in line with what the Google Books Project did. But buying and scanning may be viable for a commercial company; it is not viable for academic research. Older books that are not easily purchased increase the problem of access to the data in them. Plus, given that once a digital library is created anyone can use it, having many people buy and scan books, including hunting down older copies of books to create their own library, is inefficient. In addition, if the digital library is not secured, it could end up being shared and spread across the Internet in the same way data breaches spread other valuable data.[221] As such we propose a radical solution.

---

[219] *Accord*, Henderson et al., *supra* note 1.
[220] *See supra* notes 109 to 117 and accompanying text.

Society should create a centralized, trusted source for access to book data. The best way to do that is to have the *libraries* that have contributed to the Google Books Project Corpus (GBPC) or the Hathi Trust expand access to their data as a resource for academic research LLM training.[222] Although the Hathi Trust is a potentially viable option, we use GBP as an exemplar of how such a solution could work. As of January 27, 2023, the GBPC has "more than 40 million books in more than 500 languages"[223] It thus dwarfs the shadow libraries mere 290,000 titles and has the blessing of the law as fair use for training and non-expressive uses. Access to GBPC could be at a price per book in which case the price should not be the full price of $20 per book. The idea is not to make this approach into a cash font for the libraries and Google but instead to have a reasonable rate so research can be done not just in the U.S., but other countries given the breadth of languages in GBPC. For example, if someone wanted a corpus of 300,000 books, at a rate of $0.1 per book, that would cost $30,000. Half could go to libraries and Google (given that Google is likely the best placed to host and manage the program) and half to a pool that the Copyright Office could distribute to publishers to give to authors on a pro rata basis. We acknowledge that authors may be quite upset at the small amount of money they may receive. But this situation is not the same

[221] *Cf.* Authors Guild, Inc. v. Google, Inc. 804 F.3d 202, 228 (2d Cir. 2015) (noting with approval the security measures Google took in building and maintaining the Google Books Project).
[222] We acknowledge early critiques of the GBPC include issues around the quality of the optical character recognition scanning, metadata issues, and other technical created critiques of the corpus. *See e.g.,* Mark Davies, *Making Google Books n-grams useful for a wide range of research on language change,* 19 INTERNATIONAL JOURNAL OF CORPUS LINGUISTICS 401, 402 n.2 (2014). And yet as Google has improved just its N-Gram research offering, scholars have found the corpus useful. *Id.* at 415. Furthermore, advances in NLP research have faced messy data and so the point is better access to allow such research; not a claim that the dataset is somehow perfect.
[223] https://blog.google/products/search/google-books-library-project/#logistics

as streaming music where each time a song is played is a direct relationship to the service. Indeed, the question of how much a particular work or author contributes to an LLM output will be a challenge for plaintiffs in current lawsuits and for any payment based on amount used in an output. Even if an author or book is seen as somehow used or infringed, a given author is not likely to be more important than another in a 300,000 book corpus, let alone a larger one that may be 1 million titles.[224] There is another option that may further protect against copyright output issues.

Because the attacks on LLMs that are most troubling have yielded outputs of full chapters or other substantial amounts of a work, researchers may forgo using entire books and instead use smaller, diced up parts of books to see whether their model has learned language.[225] In this case, the repository might charge $.01 per page. Ironically, the pressure on outputs and desire to limit memorization issues may lead to less parts of books being used and so less royalties, if any, going forward. Nonetheless limiting full copies of a text addresses some problems about outputs and at a lower cost than using a full book.

Using libraries' books that are part of the GBPC as a shared repository for research has added advantages. Given that security of the book data used to train the software is an issue--a system that gave the book data to one entity but could be shared or stolen is not desired. Courts looked favorably on Google's security for its book data.[226] Google has little interest in reducing that security and is likely better placed than most entities to keep the data secure simply by keeping its practices intact. Another advantage is that you do not have to give a researcher the underlying data to use GBPC to train your model.

---

[224] This would be a hypothesis that could be tested, though at great expense. A book of gibberish words will not factor into a model the same way as a book of common idioms. The latter lends to generalization, the former not so much. One would need analytical tools such as Shapely Values. *See e.g.,* Amirata Ghorbani, , and James Zou, *Data Shapley: Equitable Valuation of Data for Machine Learning* in INTERNATIONAL CONFERENCE ON MACHINE LEARNING, pp. 2242-2251. PMLR (2019). To date, computing Shapely values---how much a unit of data impacts the overall model---requires training versions of the model with and without each unit of data. While not technically impossible, it would require basically training millions of versions of, for example, GPT3 and then comparing their outputs (there are some approximation techniques but are also prohibitively expensive at large scale). It is quite likely though, that no one unit of data in an LLM really stands out as being significantly more important than another for most books that follow typical language patterns.

[225] *See supra* Part I.B.2.

Although you might think a researcher needs a copy of the data to train their model, there are ways to let the data stay at a cloud-based service, with all the security benefits and still conduct robust training and research. As we have offered GBPC as an exemplar repository, we will stay with Google as an example of such a cloud-based service among other possible cloud-based services. There are two components needed to train a model: the code to create and train a model, and the training data. It is technically possible to send the code to construct and train a model to Google servers, where book corpus data already resides. The model is then trained on Google's servers and the external developer pays for the access to the data and for the cost of using the computing resources. In this way, Google's book data need never leave Google's computers where it could be copied. Once complete, the model can be downloaded, or even hosted permanently on Google's servers through APIs like the way OpenAI hosts ChatGPT, GPT-3, and GPT-4. Such a service could be set up to allow external developers to specify subsets of the book corpus, for example the 100,000 Yiddish titles, which will establish the payment rate. The service could also be configured to allow external developers to upload and mix in their own data. If there is an issue with the model, they can correct the code or change the data and retrain the model. This is not unlike how Google's current AutoML service already operates, which allows users to specify what kind of model they want and trains it completely within the confines of Google Cloud.[227] It is also similar to Google Colaboratory, which allows users to write arbitrary code and run it on Google's unused infrastructure, except there would be secret data that could not be downloaded.[228]

---

[226] *Cf.* Authors Guild, Inc. v. Google, Inc. 804 F.3d 202, 228 (2d Cir. 2015) (noting with approval the security measures Google took in building and maintaining the Google Books Project).
[227] https://cloud.google.com/automl?hl=en

Of course, whether libraries and Google will offer such a service is not a thing that can be mandated by law. Yet, insofar as libraries have the underlying books because they owned them, the libraries may be able to assert that this is a use they wish to enable and ask Google to aid in that offering. As a sign of true ethical behavior, if Congress passed a law with mandated fees as we describe above, Google might agree to manage the service as a public-private partnership. It may also be that libraries could see whether a Google competitor wanted to host such a service.[229] As this option may not come to fruition, we offer a simpler one.

Given that LLMs will feed the way people write and work in general and that LLMs need good inputs going forward, there could be a royalty on books sales analogous to the one in the Audio Home Recording Act (AHRA) for digital media. The music industry feared digital home recording media such as Digital Audio Tape (DAT) and recordable Compact Discs would displace income from sales of vinyl and audio cassettes because people could buy the analog copy and then make digital ones with almost loss of sound quality.[230] To address this possibility, a royalty was placed on recordable media such as digital audio tapes and recordable CDs.[231] That fee goes to the Copyright Office which then works with music publishers to distribute the royalties.[232] By analogy, part of an argument against allowing LLMs to use books is that LLMs may displace the incentives and incomes for authors and perhaps undermine the market for the high-quality writing that LLMs need and society wants.[233] Although the reality of many IP markets, especially books, is that they are winner take all markets where only a small number of authors earn enough to make a living,[234] the AHRA approach offers an acknowledgment that writing matters. With around 767 million print books sold in 2023[235] and 191 million ebooks sold in 2020[236] even a $0.50 per unit fee would generate about $479 million a year; and $1.00 would be $958 million a year. That fee combined with better control over outputs would acknowledge the importance of authorship even as authors start to embrace using LLMs to aid in writing their new works. To make sense, however, because access to books for research would still be cost-prohibitive in this option, the law would have to allow researchers to use books, even in shadow libraries, as fair use if the outputs do not infringe copyright.[237]

---

[228] https://colab.research.google.com/

[229] Although we focus on books, the idea is not limited to books. Other media such as audio recordings and video could use a repository.

[230] Joel L. McKuin, *Home Audio Taping of Copyrighted Works and the Audio Home Recording Act of 1992: A Critical Analysis* 16 HASTINGS COMM. & ENT. LJ 311, 321-322 (1993) (detailing music industry concerns over new digital recording formats).

[231] 17 U.S.C. Section 1004(d).

[232] 17 U.S.C. Sections 1006-1007.

## B. Maintaining An Open Internet

Even if we solve access to book data, access to Internet text via CommonCrawl and other sources is vital for future LLM research. Internet data is broader than book data and it is up to date regarding recent events. It also reflects, for better or for worse, how people understand history or a particular issue at a given time. Moves to close off access to data, employ restrictive terms of service, and/or use signals in code indicating "no training," "no CommonCrawl," or other limitations on access and use threaten our ability to have necessary data and indeed threaten to change the Open Internet into the sort of permission-driven situation early Internet sought to avoid.

[233] *See e.g.*, Cho, *supra* note 92.
[234] Deven R. Desai, *The Life and Death of Copyright* 2011 WISCONSIN L. REV. 219, 221-22, 257 (2011)
[235] https://www.statista.com/statistics/422595/print-book-sales-usa/
[236] https://www.statista.com/topics/1474/e-books/#topicOverview
[237] In that sense, the approach follows the logic of addressing unauthorized copying by spreading the cost and enabling access rather than trying to eliminate the practice.

Although there are research possibilities with older Internet data, they are limited and in a way decay over time. If you are trying to build an LLM model for pure language emulation, CommonCrawl from before the release of ChatGPT and its public API should be sufficient data.[238] In one sense, as long as CommonCrawl pre-ChatGPT API is available, NLP research should be viable. It may lose out on some evolution of dialects or sub-dialects that emerge in future, but as far as true NLP research, the dataset should be sufficient for the near future. Of course, as our language and how we talk about things change and as science and knowledge expand, a static dataset from 2022 will become stale and less useful. As a simple example, if the data collection were cut off before a major event, language on the Internet before that event simply could not capture how we talk about that event, *pro or con*. Imagine a dataset cut off before the summer when George Floyd was killed. A host of things, such as Black Lives Matter, antiracism, and being woke, came into our vocabulary. No matter your view of these words, the problem is they simply would not exist in the dataset with any of the richness and nuances, *including critiques*, that are part of that era until today. If the dataset were cutoff before former President Trump's first run for office, no matter your view of the existence and politics of MAGA, the problem is we would not capture all the ways language changed because of that moment in history. The question is not about whether someone likes or dislikes these events and how people spoke of them. The question is making sure an LLM can have that reality in its dataset. In that sense, this problem is more acute if your goal is to answer questions and stay up to date with results.

[238] The reason pre-ChatGPT release matters is that after its release, the Internet has changed as a test bed for language. Rather than being words assembled by humans alone, the Internet has filled up quickly with LLM generated text and that alters how well the dataset can be used for assessing whether the model assembles words as humans do. After all, if my control set has machine generated language, the set is no longer a good benchmark. *See* Melissa Heikkila, *How AI-Generated Text Is Poisoning the Internet*, MIT TECH. REV. (December 20, 2022) at https://www.technologyreview.com/2022/12/20/1065667/how-ai-generated-text-is-poisoning-the-internet/ ("In the future, it's going to get trickier and trickier to find good-quality, guaranteed AI-free training data, says Daphne Ippolito, a senior research scientist at Google Brain, the company's research unit for deep learning.") Robert McMillian, *AI Junk Is Starting to Pollute the Internet* WSJ.Com, (July 12, 2023) at https://www.wsj.com/articles/chatgpt-already-floods-some-corners-of-the-internet-with-spam-its-just-the-beginning-9c86ea25 (Should the internet increasingly fill with AI-generated content, it might become a problem for the AI companies themselves. That is because their large language models, the software that forms the basis of chatbots such as ChatGPT, train themselves on public data sets. As these data sets become increasingly filled with AI-generated content, researchers worry that the language models will become less useful, a phenomenon known as "model collapse.")

Once you move beyond language emulation to something closer to a machine that can answer all manner of questions, access to current Internet data is vital.[239] CommonCrawl and similar datasets become not only language tools but also knowledge sources. Just as cases have held that search is a transformative use of copyrighted content,[240] courts should hold that LLMs and other AI software may use such content on the Web when the use falls under fair use. The problem today is that thin caselaw and changing Internet etiquette and ethos clouds the issue.

If courts hold that ignoring restrictive terms of service or coded signals such as robots.txt, no crawl, and no training is prohibited, the nature of how we navigate the Internet and conduct research in AI will be changed for the worse. Legal scholars have deferred on whether honoring robots.txt should be part of fair use analysis,[241] noted that "disregard of robots.txt and similar mechanisms" could cut against the fourth fair use factor,[242] and suggested that "respect for technological and contractual opt-outs" should matter more for commercial rather than noncommercial research.[243] We think a stronger case is possible. Drawing on fair use doctrine, unlike books, Internet data is often freely accessible, and so using data without permission is much closer, if not the same, as when 2 Live Crew used Roy Orbison's lyrics and music to create their song.[244] Furthermore if you believe that it is proper to allow access to shadow libraries, which encompass work not intended for the open Internet, that perspective should mean you believe access to a class of copyrighted material that by design is open to access and interaction is proper regardless of signals such as no "CommonCrawl" and "robots.txt." In addition, even if something is behind a paywall, if the person accessing the data paid for it, and then used the data for data mining, software training, etc., that tracks the reverse engineering cases and so should be allowed as fair use.

[239] We acknowledge that during the course of editing this Article, efforts to create open-access, public domain datasets that avoid copyright issues. *See e.g.*, Wiggers, *supra* note 167 (describing EleutherAI and Hugging Face offerings). Although these efforts may help resolve some dataset issues, the ongoing problem of good and up-to-date data will necessarily run into copyright issues.
[240] *See* Benjamin LW Sobel. *Artificial Intelligence's Fair Use Crisis*. 41 COLUM. JL & ARTS, 45, 52-54 (2017) (discussing cases holding that copies of copyrighted images and other copyrighted material are protected as fair use)
[241] Pamela Samuelson, Christopher Jon Sprigman, and Matthew Sag, COMMENTS IN RESPONSE TO THE COPYRIGHT OFFICE'S NOTICE OF INQUIRY ON ARTIFICIAL INTELLIGENCE AND COPYRIGHT at 26 (October 30, 2023) ("It is arguable that respect for opt-outs should be part of the fair use analysis as a general consideration.")
[242] *Id*. at 24.

## Conclusion

Copyright and computer science continue to intersect and clash, but they can coexist. The advent of new technologies such as digitization of visual and aural creations, sharing technologies, search engines, social media offerings, and more challenge copyright-based industries and reopen questions about the reach of copyright law. Breakthroughs in artificial intelligence research, especially Large Language Models that leverage copyrighted material as part of training models, are the latest examples of the ongoing tension between copyright and computer science. The exuberance, rush-to-market, and edge problem cases created by a few misguided companies now raises challenges to core legal doctrines and may shift Open Internet practices for the worse. That

[243] Sag, *supra* note 104.
[244] *See supra* note 118.

result does not have to be, and should not be, the outcome. Solutions require changes from computer scientists and the copyright industry.

Computer scientists, especially those working on AI, must be careful about how their work fits into society. The quest for artificial general intelligence combined with claims about safe artificial intelligence research can delude researchers that all actions in pursuit of AGI are ok. Believing what is proper for academic research is always proper in commercial settings is another delusion. Abandoning such perspectives and instead asking, "How can we mitigate potential harms?" are better approaches. Thus, this Article presented ways to reduce the amount of a text used, examine whether access to the data was legitimate, test outputs to detect and filter results that may infringe intellectual property, and other steps to address problems with commercial uses of LLMs. In addition, the Article has noted that we are not at the dawn of the Internet. Instead, there has been an evolution and equilibrium of law balancing copyright and computer science. A body of case law around favored strategies to mitigate IP harm shows that when companies take serious steps to show respect for and effort to reduce IP harms, courts look favorably upon such steps. In contrast, when companies have used society as playground or lab, society and the law has made it clear such approaches are not allowed and carry consequences. AI-based companies should heed case law and norms rather than ignore them. At the same time, the copyright industry is not angelic either.

The copyright industry must temper its usual instinct and belief that all uses of copyrighted material should be compensated. The industry has a point about using hundreds of thousands of unauthorized copies of books is at least unethical. As we have argued that scale of use challenges current fair use doctrine and indicates such is use is not fair. Yet the drive to reject the rule that you can use copyrighted material for non-expressive, non-infringing purposes is misguided. A large problem is that buying copies of books so that a researcher can fit within fair use may be possible for commercial research, but it is out of reach for academic research. Furthermore, restricting access to books means concerns over software that had bias and under-representation in data will come to fruition. So too for Internet material. The shift to expanding website code to signal that scraping material to train software presents legal and ethical challenges. As such, this Article has offered several ways to increase access to books and Internet data. For books, we have made the case that libraries that are part of the Google Books Project could use that repository which has data in a secure environment to allow researchers to access and use that data for their research and so solve whether the book corpus was built fairly. The service would charge a fee with half of that fee going to the Copyright Office to distribute to authors. In addition, we have offered ways to compensate authors through a media subsidy as was done when other technologies threatened the music industry. In that case, the copyright industry would have to cede that access to "shadow libraries" are fair use. Similarly, if the industry wants researchers to pay for a digital copy of the book, the law should change to allow researchers to use tools to circumvent digital rights management software. For Internet material, the law should allow researchers to use openly available material regardless of code or terms of service indicating that such use is not allowed. These changes are needed, because the copyright's history of overly restrictive terms and narrowly prescribed uses of copyrighted material will only lead to bizarre clouds on and ongoing litigation over whether future software built with copyrighted material is legitimate or prohibited by contract terms.

In short, this Article has traced the technical realities of AI research, the issues differences between academic and commercial research, the ways AI research can improve its practices, and offered technical, compensatory, and legal solutions to the problems LLMs and how they are built raise. By balancing the interests of the various stakeholders and disregarding over-claims from them, the Article has shown how all concerned can behave better and compromise so that law and policy can re-establish a healthy balance between copyright and computer science.